\documentclass[12pt]{article}
 \usepackage[utf8]{inputenc}
 \usepackage[margin=1in]{geometry}
  \usepackage{setspace}
 \usepackage{amsmath}
 \usepackage{amssymb}
 \usepackage{hyperref}
 \usepackage{graphicx}
 \usepackage{caption}
 \usepackage{natbib}
 \usepackage{subcaption}
 \usepackage{color}
 \usepackage{float}

\renewcommand{\i}{i}

\newcommand{\deleq}{\stackrel{\Delta}{=}}

\newcommand{\loadep}{\delta^1}

\newcommand{\qi}{q_\i}

\newcommand{\rhothree}{\rho^3}
\newcommand{\qit}{\qi^\t}

\newcommand{\dist}{D}

\newcommand{\mi}{-\i}

\newcommand{\thetatil}{\tilde{\theta}}

\newcommand{\thetatilt}{\tilde{\theta}^{\t}}

\newcommand{\rhoone}{\rho^1}
\newcommand{\rhotwo}{\rho^2}

\newcommand{\Qmi}{Q_{-i}}

\newcommand{\Fi}{F_i}
\newcommand{\qtmi}{\qt_{\mi}}
\newcommand{\kt}{k_t}

\newcommand{\chat}{\c}

\newcommand{\Sequiv}{\S^{*}}
\newcommand{\qsran}{q^{*}}

\newcommand{\usi}{u^\s_{\i}}
\newcommand{\FP}{\Omega}

\renewcommand{\k}{t}

\newcommand{\neighinitheta}{N_{\thetaep}}
\newcommand{\neighiniload}{N_{\loadep}}
\newcommand{\neighinftheta}{N_{\thetaepfin}}
\newcommand{\neighinfload}{N_{\loadepfin}}

\newcommand{\qj}{q^j}

\newcommand{\Sbar}{\S\setminus [\thetabar]}
\newcommand{\q}{q}

\newcommand{\Q}{Q}
\newcommand{\Qi}{Q_\i}

\newcommand{\thetaep}{\epsilon^1}
\newcommand{\Uhat}{\hat{U}}
\newcommand{\loaddelta}{\delta}

\newcommand{\thetaepfin}{\bar{\epsilon}}
\newcommand{\loadepfin}{\epsilon_{\load}}

\newcommand{\qmi}{q_{-\i}}

\newcommand{\EQ}{\mathrm{EQ}}

\newcommand{\G}{G}

\newcommand{\epsilonhat}{\hat{\epsilon}}

\renewcommand{\loadepfin}{\bar{\delta}}
\renewcommand{\loaddelta}{\delta}

\newcommand{\K}{K}

\newcommand{\Costhis}{Y^{\t-1}}
\newcommand{\Loadhis}{Q^{\t-1}}

\newcommand{\qbar}{\bar{\q}}

\newcommand{\phibar}{\phi}
\newcommand{\cbar}{\c}
\newcommand{\thetasran}{\theta^{*}}

\newcommand{\sran}{s^{*}}

\renewcommand{\(}{\left(}
\renewcommand{\)}{\right)}
\renewcommand{\c}{y}

\newcommand{\phis}{\phi^\s}
\renewcommand{\t}{k}

\renewcommand{\j}{j}

\newcommand{\pro}{\mathrm{Pr}}

\renewcommand{\a}{a}
\newcommand{\thetat}{\theta^\t}

\newcommand{\cj}{\cbar^j}

\newcommand{\Ai}{A_i}
\newcommand{\ai}{a_i}
\newcommand{\actiont}{a^\t}

\newcommand{\thetazero}{\theta^1}
\newcommand{\etaj}{\log\(\frac{\phibar^{\s}(\cj|\q^j)}{\phibar^{\sran}(\cj|\q^j)}\)}

\newcommand{\etabar}{\log\(\frac{\phibar^\s(\cbar|\qbar)}{\phibar^{\sran}(\cbar|\qbar)}\)}

\newcommand{\I}{I}

\newcommand{\qhat}{\hat{\q}}
\newcommand{\qhatt}{\hat{\q}^j}

\newcommand{\epis}{\epsilon_{\i}^\s}

\newcommand{\T}{K}

\newcommand{\thetatone}{\theta^{\t+1}}

\newcommand{\Tstar}{\T^{*}}

\newcommand{\BR}{\mathrm{BR}}

\newcommand{\s}{s}

\renewcommand{\S}{S}

\newcommand{\thetabar}{\bar{\theta}}

\newcommand{\ct}{\c^\t}

\newcommand{\qt}{\q^\t}

 \usepackage{amsthm}
 \usepackage{dsfont}
 \usepackage{mathrsfs}
 \usepackage{lmodern}
 \newtheorem{assumption}{Assumption}
 \newtheorem{lemma}{Lemma}
 \newtheorem{definition}{Definition}
 \newtheorem{theorem}{Theorem}
 \newtheorem{proposition}{Proposition}

 \DeclareMathOperator*{\argmin}{arg\,min}
 
 \DeclareMathOperator*{\argmax}{arg\,max}
\newcommand{\QEDA}{\hfill $\square$}
 \usepackage{appendix}

 \title{Convergence and Stability of Coupled Belief--Strategy Learning Dynamics in Continuous Games}
 \author{Manxi Wu, Saurabh Amin, and Asuman Ozdaglar\thanks{M. Wu (manxiwu@cornell.edu) is with the School of Operations Research and Information Engineering at Cornell University. S. Amin (amins@mit.edu) is with Laboratory for Information and Decision Systems at Massachusetts Institute of Technology, and A. Ozdaglar (asuman@mit.edu) is with Department of Electrical Engineering and Computer Science at Massachusetts Institute of Technology.}}
 \date{}

\begin{document}

 \maketitle
\begin{abstract}
We propose a learning dynamics to model how strategic agents repeatedly play a continuous game while relying on an information platform to learn an unknown payoff-relevant parameter. In each time step, the platform updates a belief estimate of the parameter based on players' strategies and realized payoffs using Bayes's rule. Then, players adopt a generic learning rule to adjust their strategies based on the updated belief. We present results on the convergence of beliefs and strategies and the properties of convergent fixed points of the dynamics. We obtain sufficient and necessary conditions for the existence of globally stable fixed points. We also provide sufficient conditions for the local stability of fixed points. These results provide an approach to analyzing the long-term outcomes that arise from the interplay between Bayesian belief learning and strategy learning in games, and enable us to characterize conditions under which learning leads to a complete information equilibrium.
\end{abstract}

\section{Introduction}
Strategic agents often need to engage in repeated interactions with each other while learning an unknown environment that impacts their payoffs. Such a situation arises in online market platforms, where buyers and sellers repeatedly make their transaction decisions while learning the latent market condition that governs the price distribution. The price distribution is updated based on the previous transactions and buyer reviews on platforms such as Amazon, eBay, and Airbnb (\citet{moe2004dynamic,acemoglu2017fast}). Another situation concerns with transportation networks, where travelers make route choice decisions on a day-to-day basis while also learning the underlying state of network that affects the travel time distribution. This travel time distribution is repeatedly updated based on the delay and flow information provided by navigation apps such as Google Maps or Apple Maps (\citet{zhu2010traffic, wu2021value, meigs2017learning}). In both situations, players' strategic decisions (purchases and sales on online platforms or route choices in transportation networks) influence the learning of the unknown environment (latent market condition or network state), which then impact the players' future decisions. Thus, the long-run outcome of strategic interactions among players is governed by the joint evolution of stage-wise decisions made by the players and learning of the unknown environment.  



In this article, we introduce and study a learning model to describe this joint evolution in a game-theoretic setting. In our model, a finite number of strategic agents (players) repeatedly play a game with continuous strategy sets. The payoff of each player in the game is a function of the strategy profile with an unknown parameter (or parameter vector). A public information platform (e.g. a market platform or navigation app) serves as an information aggregator that intermittently updates and broadcasts a Bayesian belief estimate of the payoff parameter based on stage-wise game outcomes (i.e. strategies and randomly realized payoffs) to all players. Players then account for the most current belief when updating their strategies.

 The players' strategy update in our learning dynamics is generic so that it can be used to model a broad class of strategy learning rule that players decide to choose based on the updated belief and the strategy played in the previous stage. In particular, the strategy update can incorporate two classes of learning rules that naturally arise from utility-maximizing players. One is the \emph{belief-based best response dynamics}, where, in each stage, players update their strategies using the best response that maximizes their expected utility given the updated belief and their opponents' strategies. Another class is the \emph{belief-based no-regret learning dynamics}, where each player updates their strategy using an online mirror ascent algorithm that maximizes the expected utility, and the gradient used in the algorithm is estimated using the updated belief. When the true parameter is known to players, our learning dynamics reduces to the classical strategy learning in games with complete information. For this special case, the belief-based best response dynamics reduces to the best response dynamics with complete information of utility functions, and the belief-based no-regret learning reduces to the no-regret learning dynamics with complete information of the gradient of utility functions.

Past literature has extensively studied various strategy learning rules in games. Learning rules based on best response strategies have been studied in both games with finite strategy set and continuous strategy sets when players have complete information of their utility functions. This includes simultaneous and sequential best response dynamics (\citet{milgrom1990rationalizability, monderer1996potential, hofbauer2006best}), fictitious play (\citet{fudenberg1993learning, monderer1996fictitious}), and stochastic fictitious play (\citet{benaim1999mixed, hofbauer2002global}). For games with finite strategy sets, various learning rules have been proposed to model how players, who do not have complete information of the payoff matrices, learn an equilibrium based on the realized payoffs in every stage. Examples include log-linear learning (\citet{blume1993statistical}, \citet{marden2012revisiting}, \citet{alos2010logit}), regret-based learning (\citet{hart2003regret}, \citet{foster2006regret}, \citet{marden2007regret}, \citet{daskalakis2011near, syrgkanis2015fast}), payoff-based learning (\citet{cominetti2010payoff}, \citet{marden2009payoff}), and replicator dynamics (\citet{sandholm2010population}, \citet{beggs2005convergence}, \citet{hopkins2002two}). On the other hand, in games with continuous strategy sets, learning dynamics typically assume the knowledge of players' utility functions (e.g. the dynamics based on best responses), or consider that a stochastic estimate of the gradient of utility functions is available (e.g. no-regret learning and gradient-based learning in continuous games \citet{rosen1965existence, mertikopoulos2019learning, bravo2018bandit}). 

We extend the past literature by focusing on coupled belief-strategy dynamics, which naturally describes the learning behavior of players who adjust their strategies while relying on an information platform to learn the unknown utility functions that are parametrized. The setting of parametrized utility functions is especially relevant to continuous games since players need to learn their utility function on a continuous strategy set. To the best of our knowledge, past literature has not analyzed the long-run outcomes of coupled belief-strategy updates in continuous games. Our focus is to develop an approach to analyze the convergence and stability (both local and global) properties of this learning dynamics, and derive conditions under which a complete information Nash equilibrium naturally arises from the learning dynamics. Our conditions for learning complete information equilibrium can also be used for studying equilibrium computation algorithms in continuous games, however, such algorithmic aspects of equilibrium computation are outside the scope of our work.  


Next, we summarize the main contributions of our paper. 

\vspace{0.2cm}
\noindent\textbf{Convergence:} We show that the joint evolution of beliefs and strategies converges to a fixed point of our dynamics with probability 1 provided that the learning rule in strategy updates converges given a static belief (i.e. when the belief of parameter is held fixed instead of being repeatedly updated). Furthermore, at any fixed point, the fixed point belief forms a consistent estimate of the payoff distribution given the strategy, and the strategy is an equilibrium of the game corresponding to the convergent belief  (Theorem \ref{theorem:convergence}). %

Naturally, a complete information belief and the associated complete information equilibrium form a fixed point of our learning dynamics. However, the fixed point set may also include other fixed points that do not identify the true parameter. At these fixed points, the belief assigns a non-zero probability to other parameters besides the true parameter, which may lead to an incorrect payoff estimate for strategies that differ from the fixed point. Consequently, the associated fixed point strategy that the learning dynamics converge to may not be a complete information equilibrium.



Our proof of Theorem \ref{theorem:convergence} builds on the martingale property of Bayesian beliefs as well as the properties of strategy learning in games. Firstly, we obtain the convergence of Bayesian beliefs by showing that the beliefs form a martingale. Secondly, to prove the strategy convergence, we construct a series of auxiliary strategy sequences, each of which is identical to the original sequence up to a certain stage, but differs in the ``tail" subsequence -- the strategies beyond that stage are updated using the fixed point belief. We show that the distance between a so-constructed auxiliary sequence and the original sequence goes to zero as the stage beyond which the two sequences differ tends to infinity. Thus, the convergence property of strategies in our coupled belief- strategy learning dynamics is derived from the convergence of the strategy updates with static belief being the fixed point belief. Finally, using the strategy convergence result, we prove that the players' payoff distributions asymptotically approach the identical and independent distribution corresponding to the fixed point strategy. This also allows us to show that the belief concentrates exponentially fast on the subset of parameters with the property that, at fixed point, each parameter in this set induces the same payoff distribution as the true parameter.

\vspace{0.2cm}
\noindent\textbf{Global and local stability:} 
A fixed point is \emph{globally stable} if the learning dynamics starting from any initial state converges to that fixed point with probability 1. We find that globally stable fixed points exist if and only if all fixed points have complete information of the unknown parameter. Equivalently, the sufficient and necessary condition for all fixed points being complete information fixed points is that the true parameter is identifiable for any equilibrium strategy profile, i.e. no other parameter induces the same payoff distribution in equilibrium (Proposition \ref{prop:global}). In this case, learning is guaranteed to identify the true parameter, and players eventually play a complete information equilibrium. 

When the sufficient and necessary condition for global stability is not satisfied, there exist one or multiple fixed points that do not have complete information of the true parameter. Then, whether learning converges to a complete information fixed point or not depends on the initial state. In this case, we study the \emph{local stability} property that evaluates how robust a convergent fixed point is under local perturbations of beliefs and strategies. A fixed point is locally stable if states remain close to the fixed point with high probability when learning starts with an initial state close to that fixed point.  A locally stable fixed point is robust to local perturbations, and thus is more likely to arise as the long-run outcome of the learning dynamics.



We prove that a fixed point is locally stable if it satisfies two conditions (Theorem \ref{theorem:stability}): \emph{(a)} The fixed point belief is \emph{locally consistent} in that it consistently estimates the payoff distribution in a local neighborhood of the fixed point strategy (instead of just at the fixed point); \emph{(b)} The fixed point strategy has a local neighborhood that is an \emph{invariant set} for the strategy update. In proving this result, we again exploit the martingale property of beliefs and the properties of strategy updates. We use the martingale upcrossing inequality to show that under condition \emph{(a)} of local consistency, we can construct a neighborhood of initial belief, which ensures that the repeatedly updated beliefs remain in a small neighborhood of the fixed point belief with high probability. Condition \emph{(b)} further guarantees that strategy update is robust to local perturbations of strategies and beliefs. Importantly, we find that any complete information fixed point is locally stable, and thus is robust to local perturbations of belief and strategy (Proposition \ref{prop:complete_local}).  

Our global and local stability analysis contributes to the study of stability of Nash equilibrium under various learning dynamics in games. In particular, previous literature has focused on the stability of evolutionary dynamics in population games (\citet{smith1973logic, taylor1978evolutionary, samuelson1992evolutionary, matsui1992best, hofbauer2009stable, sandholm2010local}), and more recently on adaptive learning dynamics in traffic routing (\citet{dumett2018stability}). Our results extend the equilibrium stability analysis in a complete information environment to the stability analysis on both the belief estimates and players' strategies governed by the coupled belief-strategy dynamics. 


\medskip 
\noindent\textbf{Complete learning.} Learning is complete if the convergent fixed point strategy is a complete information equilibrium. Complete learning is not always guaranteed because fixed point beliefs may form incorrect estimates on the unobservable game outcomes, i.e. the payoffs of strategies that are not taken. These incorrect estimates may persist since the information of game outcomes used in belief updates is endogenously acquired based on strategies chosen by players.





The phenomenon that endogenous information acquisition leads to incomplete learning is central to a variety of settings that include multi-arm bandit problems (\citet{auer2002finite, cesa2006prediction, lattimore2020bandit}), endogenous social learning (\citet{banerjee1992simple, gale2003bayesian, golub2010naive, acemoglu2014dynamics, mossel2015strategic}), and self-confirming equilibrium learning in extensive form games (\citet{fudenberg1993self, fudenberg1995learning}). In settings where the action space of each agent is finite, complete learning can be achieved when the decision maker randomly takes each action with small probability (\citet{auer2002finite, cesa2006prediction, hart2000simple}). In our setting, players have continuous strategy sets. Thus, it is unclear under what condition exploration is needed, and how random exploration can be used to achieve complete learning. 

We focus on identifying conditions under which learning is complete without the need of exploration. We show that learning is complete if (i) the convergent fixed point assigns probability 1 to a single parameter, \emph{or} (ii) the convergent fixed point belief is locally consistent, and the payoff function of each player is concave in their strategy for any parameter (Proposition \ref{prop:equivalent_complete}). In case (i), we know that the single parameter with probability 1 is the true parameter, and the fixed point strategy is a complete information equilibrium. In case (ii), the fixed point belief does not identify the true parameter, but the local consistency condition ensures that the belief forms consistent estimate of payoffs in a local neighborhood of the fixed point strategy. Thus, each player's strategy is a local best response to their opponents' strategies. Additionally, the payoff concavity condition ensures that each player's local best response is also the true best response strategy. Therefore, in case (ii), the fixed point strategy is a complete information Nash equilibrium even though the fixed point belief does not have complete information.


If the convergent fixed point does not satisfy (i) or (ii), then learning may not be complete, and exploration of strategies other than the fixed point strategy is needed to guarantee complete learning. If payoff functions in the game are concave (which is typically assumed in most applications), then local exploration -- randomly taking strategies in a local neighborhood of the fixed point -- is sufficient to identify the complete information equilibrium. This is because local exploration can exclude any parameter that does not form locally consistent payoff estimate, and eventually leads to a locally consistent belief so that condition (ii) for complete learning is satisfied. 




The rest of the paper is organized as follows: Section \ref{sec:basic_model} describes the learning model and Section \ref{sec:main} details our main results -- convergence and stability analysis, and conditions for complete learning. We present the proofs of our main convergence and stability theorems in Section \ref{sec:proof}. In Section \ref{sec:variant}, we discuss the extensions of our results to learning with two timescales, learning in games with finite strategies, and learning with maximum a posteriori or least square estimates. We include the proofs of propositions and technical lemmas in Appendix \ref{apx:proof}. 

\color{black}

\section{Model of Learning Dynamics in Continuous Games}\label{sec:basic_model}
Our learning dynamics is induced by strategic players in a finite set $\I$ who repeatedly play a game $\G$ for an infinite number of stages. The players' payoffs in game $\G$ depend on an \emph{unknown} parameter vector $\s$ belonging to a finite set $\S$. Players have access to an information system (or an aggregator) that repeatedly updates and broadcasts a belief estimate $\theta=\(\theta(\s)\)_{\s \in \S} \in \Delta(\S)$ to all players, where $\theta(\s)$ denotes the estimated probability of parameter $\s$. 

In game $\G$, the strategy of each player $\i \in \I$ is a finite-dimensional vector $\qi$ in a convex, closed and bounded strategy set $\Qi$. The players' strategy profile is denoted $\q=\(\qi\)_{\i \in \I} \in \Q \deleq \prod_{\i \in \I} \Qi$. The payoff of each player is realized randomly according to a probability distribution. The distribution of players' payoffs $\c=\(\c_i\)_{\i\in \I}$ for any strategy profile $\q \in \Q$ and any parameter $\s \in \S$ is denoted by the probability density function $\phis(\c|\q)$. We assume that $\phis(\c|\q)$ is continuous in $\q$ for all $\s \in \S$. Without loss of generality, we write the player $\i$'s payoff $\c_{\i}$ for any $\s \in \S$ as the sum of an average payoff function $\usi(\q)$ that is continuous in $\q$ and a zero-mean noise term $\epis(\q)$ that can be correlated across players:
\begin{align}\label{eq:utility}
    \c_i=\usi(\q)+\epis(\q).
\end{align}


We model our learning dynamics as a discrete-time stochastic process, with state comprising of belief estimate of the unknown parameter and the players' strategies: In each stage $\t \in \mathbb{N}_{+}$, the information platform broadcasts the current belief estimate $\thetat$; the players act according to a strategy profile $\qt=\(\qt_i\)_{\i \in \I}$; and the payoffs $\ct=\(\ct_i\)_{\i \in \I}$ are realized according to $\phi^s(\ct|\qt)$ when the parameter is $\s \in \S$. The state of the learning dynamics in stage $\t$ is $\(\thetat, \qt\) \in \Delta(\S) \times \Q$. 

The unknown true parameter is $\sran \in \S$. We suppose that the initial belief $\thetazero$ does not exclude any possible parameter, i.e. $\thetazero(\s)>0$ for all $\s \in \S$, and the initial strategy $\q^1 \in \Q$ is feasible. The evolution of states $\(\thetat, \qt\)_{\t=1}^{\infty}$ is jointly governed by belief and strategy updates as introduced next.

\vspace{0.2cm}
\noindent\textbf{Belief update.} The information platform updates the belief intermittently and infinitely. The stages at which the belief is updated can be deterministic or random, denoted by the subsequence $\(\kt\)_{\k=1}^{\infty}$. In stage~$\t_{t+1}$, the most recent belief estimate $\theta^{\kt}$ is updated using players' strategy profiles $\(\qt\)_{\t=\kt}^{\t_{t+1}-1}$ and payoffs $\(\ct\)_{\t=\kt}^{\t_{t+1}-1}$ between the stages $\t_{t}$ and $\t_{t+1}$ according to the Bayes' rule:\footnote{In many instances of the problem setup, it is sufficient to update the belief only based on aggregate strategies $\tilde{\q}^\t$ and payoffs $\tilde{\c}^\t$, provided that the tuple $\(\tilde{\q}^\t, \tilde{\c}^\t\)$ is a sufficient statistics of $\(\qt, \ct\)$ in the following sense: one can write $\phibar^\s(\ct|\qt)= \psi\(\qt, \ct|\tilde{\q}^\t, \tilde{\c}^\t\) \tilde{\phi}^\s\(\tilde{\c}^\t|\tilde{\q}^\t\)$, where $\psi\(\qt, \ct|\tilde{\q}^\t, \tilde{\c}^\t\)$ is the $\s$-independent conditional distribution of strategy and payoffs given the aggregate statistics, and $\tilde{\phi}^\s\(\tilde{\c}^\t|\tilde{\q}^\t\)$ is the conditional probability of $\tilde{\c}^\t$ given $\tilde{\q}^\t$ for parameter $\s$. Then, we can re-write \eqref{eq:update_belief} as Bayesian update that only relies on $\(\tilde{\q}^\t, \tilde{\c}^\t\)$: 
\begin{align*}
    \theta^{k_{t+1}}(\s)&= \frac{\theta^{\kt}(\s)\prod_{\t=\kt}^{\t_{t+1}-1}\psi(\qt, \ct|\tilde{\q}^\t, \tilde{\c}^\t) \tilde{\phi}^s(\tilde{\c}^\t|\tilde{\q}^\t)}{\sum_{s' \in \S} \theta^{\kt}(\s') \prod_{\t=\kt}^{\t_{t+1}-1}\psi(\qt, \ct|\tilde{\q}^\t, \tilde{\c}^\t) \tilde{\phi}^{s'}(\tilde{\c}^\t|\tilde{\q}^\t)}= \frac{\theta^{\kt}(\s) \prod_{\t=\kt}^{\t_{t+1}-1}\tilde{\phi}^s(\tilde{\c}^\t|\tilde{\q}^\t)}{\sum_{s' \in \S} \theta^{\kt}(\s') \prod_{\t=\kt}^{\t_{t+1}-1} \tilde{\phi}^{s'}(\tilde{\c}^\t|\tilde{\q}^\t)}, \quad \forall \s \in \S. 
\end{align*}
For example, in routing games, the aggregate traffic flow and travel time of each edge in the network are sufficient statistics of the individual travelers' route choices and travel time. Another example is in Cournot games, the total production level and product price in a market are sufficient statistics of each producer's production and revenue (see Section \ref{subsec:example}).}
\begin{align}
    \theta^{k_{t+1}}(\s)&= \frac{\theta^{\kt}(\s)\prod_{\t=\kt}^{\t_{t+1}-1}\phibar^\s(\ct|\qt)}{\sum_{s' \in \S} \theta^{\kt}(\s') \prod_{\t=\kt}^{\t_{t+1}-1}\phibar^{\s'}(\ct|\qt)}, \quad \forall \s \in \S. \tag{$\theta$-update} \label{eq:update_belief}
\end{align}

\vspace{0.2cm}
\noindent \textbf{Strategy update.} Players update their strategies in each stage based on the updated belief and the current strategies played by their opponents. Given any $\thetatone$ and any $\qt$, we denote the strategy update for each $\i \in \I$ as a set-valued function $\Fi\(\thetatone, \qt\): \Delta\(\S\) \times \Q \rightrightarrows \Qi$: 
\begin{align}\tag{$\q$-update}\label{eq:generic}
    \qi^{\t+1} \in \Fi\(\thetatone, \qt\), \quad \forall \i \in \I.
\end{align}

In our model, the belief and strategy updates \eqref{eq:update_belief} -- \eqref{eq:generic} capture the dynamic interplay between Bayesian parameter learning and strategy learning in games.
Specifically, since the distribution of $\ct$ depends on the strategy profile $\qt$ in each stage, the sequence of payoffs $(\ct)_{\t=1}^{\infty}$ is not identically and independently distributed when the strategies $(\qt)_{\t=1}^{\infty}$ are repeatedly updated. On the other hand, the strategy updates \eqref{eq:generic} depend on the repeatedly updated beliefs $(\thetat)_{\t=1}^{\infty}$. 

We note that the belief updates can occur less frequently than the strategy updates since the subsequence of belief update stages satisfy $\t_{t+1}- \kt \geq 1$. In Sec. \ref{sec:main} -- \ref{sec:proof}, we present our main results by considering that $\t_{t+1}- \kt$ is finite with probability (w.p.) 1; i.e., both belief and strategy updates follow the same timescale. In Sec. \ref{sec:variant}, we show that all results analogously hold when $\lim_{t \to \infty} \t_{t+1}- \kt=\infty$ w.p.1; that is when the belief is updated at a slower timescale compared to the strategy updates. 


For any belief $\theta$, we denote $G(\theta)$ as the game with a static information environment, where each player $i$'s utility function is the expected utility $\mathbb{E}_{\theta}\left[u_i^s(\qi, \qmi)\right] = \sum_{\s \in \S} \theta(\s)u_i^s(\qi, \qmi)$. Note that if the belief $\thetat$ were held constant as $\theta$ for all $\t$, then \eqref{eq:generic} simply becomes the corresponding strategy update $q^{\t+1} = \Fi(\theta, \qt)$ in game $G(\theta)$. Since our goal is to analyze the joint evolution of both the beliefs and the strategies (instead of the properties of a particular strategy update in static information environment), we make the following two assumptions throughout the paper:


\begin{assumption}[Upper-hemicontinuous strategy updates]\label{as:uh}
For any $i \in I$, $\Fi(\theta, \q)$ is upper hemicontinuous in $\theta$ and $\q$. 
\end{assumption}

\begin{assumption}[Convergence in static information environment]\label{as:convergence}
For any belief $\theta$, the equilibrium strategy set of the game $G(\theta)$ is non-empty. Moreover, given any initial strategy, the strategy update \eqref{eq:generic} with static belief $\theta$ converges to an equilibrium in game $G(\theta)$. 
\end{assumption}\begin{assumption}\label{as:isolation}
For any $\theta \in \Delta(S)$, either (i) or (ii) is satisfied: 
\begin{itemize}
    \item[(i)] $\EQ(\theta)$ is a singleton set. 
    \item[(ii)] There exists $\delta>0$ such that any two different equilibria $q^{\dagger}, q^{\ddagger} \in \EQ(\theta)$ satisfy $\|q^{\dagger} - q^{\ddagger}\|> \delta$. Moreover, for any $q^{\dagger} \in \EQ(\theta)$, there exists $\xi_{a}, \xi_{b}>0$ such that for any $q$ that satisfies $\|q- q^{\dagger}\|< \xi_a$ and any $\theta$ that satisfies $\|\theta - \thetabar\|< \xi_b$, the strategy update $F(\theta, q) = (F_i(\theta, q))_{i \in I}$ is unique. 
\end{itemize}
\end{assumption}

Assumption \ref{as:uh} ensures that the updated strategy is upper hemicontinuous in the belief and the strategy in the previous stage. Assumption \ref{as:convergence} guarantees the convergence of strategies induced by updates under static information environment. Without Assumption \ref{as:convergence}, strategies may not converge even in the static information environment, and consequently the joint evolution of both the beliefs and the strategies also cannot be guaranteed to converge. {Assumption \ref{as:isolation} implies that when the game has multiple equilibria, each equilibrium is isolated, and the strategy update has a unique value in the local neighborhood of each equilibrium. }

%


Our results and analysis approach hold for all types of strategy update \eqref{eq:generic} that satisfy Assumptions \ref{as:uh} -- \ref{as:isolation}. For the sake of concreteness, we provide two examples of $\Fi$ that are natural extension of widely studied strategy update rules: 
\medskip 
\begin{enumerate}
    \item \emph{Belief-based best response dynamics.} We define the best response correspondence of each player $\BR_i(\thetatone, \qtmi) \deleq \argmax_{\qi \in \Qi} \sum_{\s \in \S} \thetatone(\s)u_i^s(\qi, \qtmi)$ as the set of strategies that maximize their expected utility given the opponents' strategies $\qtmi$ and the updated belief $\thetatone$. The strategy is updated following any one of the following dynamics: \eqref{eq:sbr} -- each player chooses a best response strategy; \eqref{eq:br} -- players sequentially update their strategies as the best response; \eqref{eq:br_diminish} -- each player updates their strategy as a linear combination of their current strategy and a best response strategy.  
    \begin{align}
        \Fi(\thetatone, \qt)&= \BR_i(\thetatone, \qtmi), \quad \forall \i \in \I, \quad \forall k. \tag{Simultaneous-BR}\label{eq:sbr}\\
         \Fi(\thetatone, \qt)&=\left\{\begin{array}{ll}  \BR_i(\thetatone, \qtmi), &\quad \text{if } \t ~ \text{mod} ~|\I| = \i,  \\
        \qi^{\t},& \quad \text{otherwise}. 
        \end{array}
        \right.\tag{Sequential-BR}\label{eq:br}\\
        \Fi(\thetatone, \qt)&=(1-\alpha^k)\qit+ \alpha^{\t} \BR_i(\thetatone, \qt_{\mi}), \quad \forall \i \in \I, \quad \forall \t, \tag{Inertial-BR}\label{eq:br_diminish} 
    \end{align}
 
       
where $\alpha^\t \in [0, 1]$ in \eqref{eq:br_diminish} for each $\t$.
\item \emph{Belief-based no-regret learning.} Given the updated belief $\thetatone$, and the current strategy profile $\qt$, the updated strategy $\q^{k+1}$ is given by: 
\begin{align}\label{eq:gradient}\tag{No-regret}
   \q_i^{k+1} = \argmax_{\qi \in \Qi} \{\left(x_i^{k+1}\right)^T \cdot \qi - h_i(\qi)\}, \quad \forall i \in I, \quad \forall k,
   \end{align}
   where the function $h_i:\Qi \to \mathbb{R}$ is a regularizer that is continuous and strongly convex in $\qi$\footnote{The function $h$ is strongly convex if there exists $A>0$ such that $h(\lambda q + (1 - \lambda)q') \leq \lambda h(q) + (1 - \lambda)h(q') - (1/2)A\lambda (1-\lambda)\|q-q'\|^2$ for all $q, q' \in Q$ and all $\lambda \in  [0, 1]$}, and $(x_i^k)_{k=1}^{\infty}$ is a sequence of each player $i$'s scores such that $x_i^1=q_i^1$, and 
   \begin{align*}
   x_{i}^{k+1} = x_i^k + \alpha^k \left(\sum_{s \in S} \thetatone(s) \frac{\partial u_i^s(\qt)}{\partial \qit}\right), \quad \forall i \in I, \quad \forall k.
\end{align*}
That is, the initial score of player $i$ is the same with the strategy $q^1_i$, and the score in each step is updated using gradient ascent without  regard to the feasibility constraint imposed by strategy set $Q_i$. In particular, the  gradient vector is the gradient of the expected utility function $\partial \mathbb{E}_{\theta^{k+1}}[u_i^s(q^k)]/\partial q_i^k = \sum_{s \in S} \thetatone(s) \frac{\partial u_i^s(\qt)}{\partial \qit}$ based on the updated belief $\thetatone$.\footnote{Our belief-based no-regret learning estimates the gradient of the expected utility function $\mathbb{E}_{\theta}[u_i^s(q)]$ based on the belief $\theta$ and the parametrized utility function $u_i^s(q)$. When the belief $\theta$ does not have complete information of the true parameter $\sran$, the estimated gradient is different from the gradient of the true utility function $u_i^{\sran}(q)$. As the belief converges, the estimated gradient becomes unbiased of the true gradient.}

Then, the updated strategy $q^{k+1}$ is a feasible strategy profile in $Q$ that minimizes the inner product between the strategy and the updated score vector $x^{k+1}=(x_i^{k+1})_{i \in I}$ plus the value of the regularizer. A common example of such regularizer is $h_i(\qi)=(1/2)\|\qi\|^2$. In this case, the updated strategy is the projection of the score vector $x^{k+1}$ onto the feasible strategy set; i.e. $q_i^{k+1} = \argmin_{q_i\in Q_i} \|q_i - x_i^{k+1}\|^2$. Thus, the strategy update follows a projected gradient ascent algorithm based on the updated belief for each player. 
\end{enumerate}

We note that Assumption \ref{as:uh} is satisfied by the belief-based best response learning and the belief-based no-regret learning when the utility function $u_i^s(\q)$ is continuously differentiable with respect to $\q$ for all $i\in I$ and all $\s \in \S$.\footnote{Following the Berge's maximum theorem, for any $\theta \in \Delta(\S)$, any $\i \in \I$ and any $\qmi \in \Qmi$, $\BR(\theta, \qmi)$ is upper-hemicontinuous in $\theta$ and $\qmi$. Therefore, the three types of belief-based best response updates satisfy Assumption \ref{as:uh}. Additionally, when the utility function $u_i^s(\q)$ is continuously differentiable with respect to $\q$ for any $s \in S$ and any 
$i \in I$, the updated belief given by \eqref{eq:gradient} is also upper-hemicontinuous in $\theta$ and $\q$.} From the extensive literature in learning in games (\citet{fudenberg1993learning, sandholm2010population}), we know that convergence of strategy updates to equilibrium is not guaranteed in arbitrary games even under static information environment. In Table \ref{tab:summary}, we provide a list of games in which the strategy updates with the two examples of learning rules satisfy Assumption \ref{as:convergence}. Our results, as discussed next, hold for any \eqref{eq:generic} satisfying the two assumptions.

\begin{table}[htp]
    \centering
    \begin{tabular}{p{4cm}|p{11.5cm}}
    \hline
    $F(\theta, q)$ & Game $G(\theta)$\\
    \hline
        \eqref{eq:sbr} & Two-player Cournot games, and Dominance solvable supermodular games (\citet{monderer1996potential, milgrom1990rationalizability}) \\
         \hline
        \eqref{eq:br} & Potential games (\citet{monderer1996potential})\\
         \hline
         \eqref{eq:br_diminish} &  Potential games, convex-concave zero-sum games, dominance solvable games, and supermodular games (\citet{monderer1996fictitious, hofbauer2006best, milgrom1990rationalizability})\\
         \hline 
         \eqref{eq:gradient} & Concave potential games, and strictly concave games (\citet{rosen1965existence, mertikopoulos2019learning, golowich2020tight}) \\
         \hline
    \end{tabular}
    \caption{Examples of games in which belief-based best response learning and belief-based no-regret learning satisfy Assumption \ref{as:convergence}.}
    \label{tab:summary}
\end{table}

\color{black}
\section{Main Results}\label{sec:main}
Sec. \ref{subsec:convergence} presents the convergence of beliefs and strategies. Sec. \ref{subsec:stability} analyzes the local and global stability properties of fixed points. Sec. \ref{subsec:complete} focuses on conditions for learning to converge to equilibrium with complete information, and Sec. \ref{subsec:example} illustrates our results through examples. We focus on presenting the results and explaining the intuition in this section. We defer the proofs of Theorems to Sec. \ref{sec:proof}, and the proofs of Propositions to Appendix \ref{apx:proof}.
\subsection{Convergence}\label{subsec:convergence}
To present our convergence result, we need to introduce two definitions.
\begin{definition}[Kullback–Leibler (KL)-divergence]\label{def:KL}
For a strategy profile $\q\in\Q$, the KL divergence between the payoff distribution with parameters $\s$  and $\sran\in\S$ is:
\begin{align*}
     D_{KL} \left(\phibar^{\sran}(\cbar|\q)
     ||\phibar^{\s}(\cbar|\q)\right) \deleq \left\{
     \begin{array}{ll}
     \int_{\chat} \phibar^{\sran}(\chat|\q) \log\left(\frac{\phibar^{\sran}(\chat|\q)}{\phibar^{\s}(\chat|\q)}\right) d\chat, & \quad \text{if $\phibar^{\sran}(\chat|\q) \ll \phibar^{\s}(\chat|\q)$}, \\
     \infty & \quad \text{otherwise.}
     \end{array}
     \right.
\end{align*}
\end{definition}
Here $\phibar^{\sran}(\chat|\q) \ll \phibar^{\s}(\chat|\q)$ means that the distribution $\phibar^{\sran}(\chat|\q)$ is absolutely continuous with respect to $\phibar^{\s}(\chat|\q)$, i.e. $\phibar^{\s}(\chat|\q)=0$ implies $\phibar^{\sran}(\chat|\q)=0$ w.p. 1.

\begin{definition}[Payoff-equivalent parameters]\label{def:payoff_equivalence}
A parameter $\s\in\S$ is payoff-equivalent to the true parameter $\sran$ for a strategy $\q\in\Q$ if $D_{KL} \left(\phibar^{\sran}(\cbar|\q)||\phibar^{\s}(\cbar|\q)\right)=0$. For a given strategy profile $\q\in Q$, the set of parameters that are payoff-equivalent to $\sran$ is:  
\begin{align*}
    \Sequiv(\q)\deleq \{s \in \S|D_{KL} \left(\phibar^{\sran}(\cbar|\q)||\phibar^{\s}(\cbar|\q)\right)=0\}.
\end{align*}
\end{definition}

The KL-divergence between any two distributions is non-negative, and is equal to zero if and only if the two distributions are identical. For a given strategy profile $\q \in \Q$, if a parameter $\s$ is in the payoff-equivalent parameter set $\Sequiv(\q)$, then the payoff distribution is identical for parameters $\s$ and $\sran$, i.e. $\phibar^{\sran}(\chat|\q)=\phibar^{\s}(\chat|\q)$ for all $\chat$. In this case, realized payoffs cannot be used to distinguish $\s$ and $\sran$ in the belief update \eqref{eq:update_belief} because the belief ratio $\frac{\thetat(\s)}{\thetat(\sran)}$ remains unchanged w.p. 1. Also note that the set $\Sequiv(\q)$ can vary with strategy profile $\q$, and hence a payoff-equivalent parameter for one strategy profile may not be payoff-equivalent for another strategy profile.



\begin{theorem}\label{theorem:convergence}
For any initial state~$(\theta^1, \q^1)\in\Delta(S)\times \Q$, under Assumptions \ref{as:uh} -- \ref{as:isolation}, the sequence of states $(\thetat, \qt)_{\t=1}^{\infty}$ induced by \eqref{eq:update_belief} and \eqref{eq:generic} converges to a fixed point $(\thetabar, \qbar)$ w.p. 1, and $\(\thetabar, \qbar\)$ satisfies: 
\begin{subequations}\label{eq:fixed_point_def}
\begin{align}
    [\thetabar] &\subseteq \Sequiv(\qbar), \label{eq:exclude_distinguished}\\
    \qbar&\in \EQ(\thetabar), \label{subeq:eq_fixed}
\end{align}
\end{subequations}
where $[\thetabar] \deleq \{s \in \S|\thetabar(\s)>0\}$, and $\EQ(\thetabar)$ is the set of equilibria corresponding to belief $\thetabar$. 

Moreover, for any $\s \in \S\setminus \Sequiv(\qbar)$, if $\phibar^{\sran}(\chat|\qbar) \ll \phibar^{\s}(\chat|\qbar)$, then $\thetat(\s)$ converges to 0 exponentially fast: 
\begin{align}
    \lim_{\t \to \infty} \frac{1}{\t} \log(\thetat(\s))=-D_{KL}(\phibar^{\sran}(\cbar|\qbar)||\phibar^{\s}(\cbar|\qbar)), \quad  w.p.~1.\label{eq:rate}
\end{align}
Otherwise, there exists a positive integer $\Tstar<\infty$ such that $\thetat(\s)=0$ for all $\t>\Tstar$ w.p. 1. 
\end{theorem}


This result demonstrates two facts: Firstly, if a strategy update converges in static information environment, then this strategy update also converges when beliefs are repeatedly updated. That is, the convergence property of strategy update dynamics in games with static information environment also holds under joint evolution of belief learning and strategy learning. 

Secondly, \eqref{eq:exclude_distinguished} ensures that at any fixed point $\(\thetabar, \qbar\)$, the payoff distribution $\mu(\chat|\thetabar, \qbar)$ estimated by the fixed point belief $\thetabar$ is \emph{consistent} with the true distribution given $\qbar$, i.e.  
\begin{align}\label{eq:marginal}
    \mu(\chat|\thetabar, \qbar)\deleq\sum_{\s \in \S} \thetabar(\s) \phibar^\s(\chat|\qbar)\stackrel{\eqref{eq:exclude_distinguished}}{=}\sum_{\s \in \Sequiv(\qbar)}\thetabar(\s) \phibar^{\s}(\cbar|\qbar)=\sum_{\s \in \Sequiv(\qbar)} \thetabar(\s) \phibar^{\sran}(\cbar|\qbar)= \phibar^{\sran}(\cbar|\qbar), \quad \forall \c. 
 \end{align}
Belief consistency follows from the fact that any parameter that is not payoff equivalent to $\sran$ is excluded from the support set of $\thetabar$. In fact, we show in \eqref{eq:rate} that the belief of such parameter $\s \in \S\setminus \Sequiv(\qbar)$ converges to zero exponentially fast with the rate of convergence being the KL-divergence between the corresponding payoff distribution with $s$ and the true parameter $\sran$ given strategy $\qbar$. 

Moreover, \eqref{subeq:eq_fixed} ensures that $\qbar$ is an equilibrium with respect to the estimated payoff distribution. Therefore, at a fixed point, players have no incentive to deviate from $\qbar$, and the realized payoffs can no longer change the belief $\thetabar$.

We define the set of all fixed points as $
    \FP \deleq \left\{\(\thetabar, \qbar\)\left\vert [\theta] \subseteq \Sequiv\(\qbar\), ~ \qbar \in \EQ(\thetabar)\right.\right\}$. We denote the belief vector $\thetasran$ with $\thetasran(\sran)=1$ as the \emph{complete information belief}, and any strategy $\qsran\in \EQ(\thetasran)$ as a \emph{complete information equilibrium}. Since $[\thetasran]=\{\sran\} \subseteq \Sequiv(\qsran)$, the state $\(\thetasran, \qsran\)$ is always a fixed point (i.e. $\(\thetasran, \qsran\) \in \FP$), and has the property that all players have complete information of the true parameter $\sran$ and choose a complete information equilibrium. Therefore, we refer to $\(\thetasran, \qsran\)$ as a \emph{complete information fixed point}.

However, the set $\FP$ may contain other fixed points $\(\thetabar, \qbar\)$ that are not equivalent to the complete information environment (i.e. $\thetabar \neq \thetasran$). Such belief $\thetabar$ must assign positive probability to at least one parameter $\s \neq \sran$ that is payoff-equivalent to $\sran$ given the fixed point strategy profile $\qbar$, but not necessarily at all $\q \in \Q$. We refer to such fixed points as incomplete information fixed points. 

Importantly, incomplete information fixed points arise in our learning dynamics due to the fact that realized payoffs are governed by the distribution corresponding to $\qbar$ at fixed point, which may not distinguish every other parameter $s$ from the true parameter $\sran$. Consequently, the fixed point strategy $\qbar$ may not be a complete information equilibrium. In Sec. \ref{subsec:complete}, we analyze the difference between complete information fixed points and incomplete information fixed points. We also provide conditions under which learning leads to complete information fixed points with probability 1. 


\subsection{Local and Global Stability}\label{subsec:stability}

We now analyze the global and local stability of fixed points. The definitions of global and local stability are as follows: 
\begin{definition}[Global stability]\label{def:global}
A fixed point belief $\thetabar \in \Delta(\S)$ and the associated equilibrium set $\EQ(\thetabar)$ are \emph{globally stable} if for any initial state $\(\thetazero, \q^1\)$, the beliefs of the learning dynamics $\(\thetat\)_{\t=1}^{\infty}$ converge to $\thetabar$ and the strategies $\(\qt\)_{\t=1}^{\infty}$ converge to $\EQ(\thetabar)$ with probability 1. 
\end{definition}

The definition of global stability requires that that the convergent fixed point does not depend on the initial state of the learning dynamics. On the other hand, local stability requires that the learning dynamics is robust to small perturbations around $\thetabar$ and $\EQ(\bar\theta)$. Such local perturbations of beliefs may arise from random errors in data collection or analysis algorithms, and local perturbations of strategies are likely to occur due to players' mistakes in choosing strategies or random local exploration. 
\begin{definition}[Local stability]\label{def:local}
A fixed point belief $\thetabar \in \Delta(\S)$ and the associated equilibrium set $\EQ(\thetabar)$ are \emph{locally stable} if for any $\gamma \in (0,1)$ and any $\thetaepfin, \loadepfin>0$, there exist $\thetaep, \loadep>0$ such that for the learning dynamics \eqref{eq:update_belief} and \eqref{eq:generic} starting from $\thetazero\in \neighinitheta(\thetabar)$ and $\q^1 \in \neighiniload(\EQ(\thetabar))$, the following holds:
\begin{align}\label{eq:local}
\lim_{\t \to \infty} \pro\(\thetat \in \neighinftheta(\thetabar), ~ \qt\in \neighinfload(\EQ(\thetabar))\)> \gamma.\end{align}
\end{definition}

Note that both global and local stability notions are not defined for a single fixed point, but rather for the
tuple $\(\bar\theta, \EQ(\bar\theta)\)$, i.e. fixed points with an identical belief $\bar\theta$. This is important when the game has multiple equilibria; i.e., $\EQ(\theta)$ is not a singleton set for some belief $\theta\in\Delta(S)$. That is, our stability notions do not hinge on the convergence to a particular equilibrium in the fixed point equilibrium set $\EQ(\thetabar)$.


The next proposition provides two conditions that are equivalent to the existence of globally stable fixed points, and shows that any globally stable fixed point must have complete information.


\begin{proposition}\label{prop:global}
The following three statements are equivalent: 
\begin{enumerate}
    \item[(a)] The set of globally stable fixed points is non-empty. 
 \item[(b)] For any $\theta \in \Delta(\S) \setminus \{\thetasran\}$ and any $q \in \EQ(\theta)$, there exists $\s \in [\theta]$ such that $ \s \notin \Sequiv(\q)$.
    \item[(c)] All fixed points are complete information fixed points, i.e. $\Omega = \left\{\(\thetasran, \EQ(\thetasran)\)\right\}$. 
\end{enumerate}
Moreover, any globally stable fixed point must be a complete information fixed point. 
\end{proposition}
In Proposition \ref{prop:global}, condition \emph{(a)} states the existence of globally stable fixed point; Condition \emph{(b)} states that any belief that does not have complete information cannot be a fixed point belief, because $\sran$ can be distinguished at an associated equilibrium strategy (i.e. \eqref{eq:exclude_distinguished} is not satisfied); Condition \emph{(c)} states that all fixed points are complete information fixed points. 

Proposition \ref{prop:global} is intuitive: Clearly, \emph{(b)} -- states that any belief that does not have complete information cannot be a fixed point belief -- is equivalent to \emph{(c)} -- all fixed points have complete information. Additionally, \emph{(a)} is equivalent to \emph{(c)} because if the fixed point set $\FP$ contains another fixed point that is not a complete information fixed point, then whether the states of learning dynamics converge to the complete information fixed point or another fixed point depends on the initial state; hence no fixed point in such a set can be globally stable. On the other hand, when all fixed points are complete information fixed points, states of the learning dynamics must converge to $\thetasran$ and the associated equilibrium set $\EQ(\thetasran)$ regardless of the initial state, and thus must be globally stable. Consequently, \emph{(a)} and \emph{(b)} must also be equivalent. Given the equivalence between \emph{(a)} and \emph{(c)}, it directly follows that all globally stable fixed points must be complete information fixed points. 



From Proposition \ref{prop:global}, we know that if there exist fixed points with incomplete information, then no fixed point is globally stable. In other words, whether learning converges to a complete information fixed point or a fixed point with incomplete information depends on the initial state. How likely will a particular fixed point arise as the long-run outcome of our learning dynamics? We answer this question by analyzing the local stability property of fixed points. In Theorem \ref{theorem:stability}, we provide sufficient conditions for local stability of fixed point, i.e. if learning starts with state close to a fixed point that satisfies these conditions, then states will remain close to that fixed point with high probability. We further show in Proposition \ref{prop:complete_local} that all complete information fixed points are locally stable.

Before proceeding, for any $\epsilon>0$, we define an $\epsilon$-neighborhood of belief $\thetabar$ as $N_{\epsilon}(\thetabar) \deleq \left\{\theta \left\vert \|\theta - \thetabar\| < \epsilon\right. \right\}$, where $\|\cdot\|$ is the Euclidean distance. For any $\delta>0$, we define the $\delta$-neighborhood of equilibrium set as $N_{\delta}(\EQ(\thetabar)) \deleq  \left\{ \q \in \Q\left\vert \dist\(\q, \EQ(\thetabar)\)< \delta\right.\right\}$, where $\dist\(\q, \EQ(\thetabar)\) = \min_{\q'\in \EQ(\thetabar)}\|\q-\q'\|$ is the Euclidean distance between $\q$ and the equilibrium set $\EQ(\thetabar)$.\footnote{Generically, the function $\dist\(\q, \hat{Q}\) = \min_{\q'\in \hat{Q}}\|\q-\q'\|$ defines the Euclidean distance between $\q$ and any strategy subset $\hat{Q}$.}  We introduce the following assumption: 
\begin{assumption}\label{as:stability}
For a fixed point belief $\thetabar$ and the associated equilibrium set $\EQ(\thetabar)$, $\exists \epsilon, ~\delta >0$ such that the neighborhoods $N_{\epsilon}\(\thetabar\)$ and $N_{\loaddelta}\(\EQ(\thetabar)\)$ satisfy\\
(A3a)  \emph{Local consistency}: Fixed point belief $\thetabar$ forms a consistent payoff estimate in the local neighborhood $N_\delta(EQ(\bar\theta))$, that is $[\bar\theta]\subseteq \Sequiv(q)$ for any $q\in N_\delta(EQ(\bar\theta))$. \\
(A3b) \emph{Local invariance}: Neighborhood $N_{\loaddelta}(\EQ(\thetabar))$ is a locally invariant set of the strategy update, that is, $F(\theta, \q) \subseteq N_{\loaddelta}(\EQ(\thetabar))$ for any  $\theta \in N_{\epsilon}\(\thetabar\)$ and any $\q \in N_{\loaddelta}\(\EQ(\thetabar)\)$.
\end{assumption}

\begin{theorem}\label{theorem:stability}
Let the learning dynamics \eqref{eq:update_belief} and \eqref{eq:generic} satisfy Assumptions \ref{as:uh} -- \ref{as:isolation}. Then, a fixed point belief $\thetabar$ and the associated equilibrium set $EQ(\bar\theta)$ is locally stable if Assumption \ref{as:stability} is satisfied. 
\end{theorem}







From Theorem \ref{theorem:convergence}, we already know that Assumptions \ref{as:uh} -- \ref{as:isolation} guarantee the convergence of beliefs and strategies under local perturbations. We now discuss the role of Assumption \ref{as:stability} towards ensuring local stability of a fixed point. 
Firstly, the local consistency condition \emph{(A3a)} ensures that \eqref{eq:update_belief} keeps the beliefs close to $\thetabar$. In particular, under this condition, any parameter in the support of $\thetabar$ remains to be payoff equivalent to $\sran$ for any strategy in a local neighborhood of $\EQ(\thetabar)$. Then, $\thetabar$ forms a consistent estimate of players' payoffs not just at fixed point strategy $\qbar$, but also when the strategy is locally perturbed around $\qbar$. Therefore, the Bayesian belief update keeps the beliefs of all parameters in $[\thetabar]$ close to their respective probabilities in $\thetabar$ when the strategies are in the local neighborhood, and eventually any parameters not in $[\thetabar]$ are excluded by the learning dynamics. 

Secondly, the local invariance condition \emph{(A3b)} guarantees that the strategy sequence resulting from the strategy updates remains within the locally invariant neighborhood of the fixed point equilibrium. Due to the dynamic interplay between the belief updates and strategy updates, we need both local consistency and local invariance conditions to ensure that the strategy sequence in our learning dynamics does not leave the local neighborhood of $\EQ(\thetabar)$ and the perturbed beliefs remain close to $\thetabar$.

Finally, we demonstrate that any complete information fixed point must be locally stable. 
\begin{proposition}\label{prop:complete_local}
Any complete information fixed point $(\thetasran, \qsran)$ is locally stable. 
\end{proposition}
This result implies that any complete information fixed point is robust to local perturbations of beliefs and strategies. It directly follows from Theorem \ref{theorem:stability} since any $(\thetasran, \qsran)$ can be shown to satisfy Assumption \ref{as:stability}. To see this, note that the complete information belief $\thetasran$ satisfies local consistency since it forms a consistent payoff estimate for any feasible strategy $q \in Q$ (and thus for any local neighborhood of $\qbar$). Besides, the local invariance condition is satisfied because the set $Q$ is an invariant set for any belief $\theta$ (and thus for any belief in a local neighborhood of $\thetabar$). 

On the other hand, other fixed points $(\thetabar, \qbar)$  are not guaranteed to be locally stable unless there exists a local neighborhood of it that satisfies local consistency and local invariance. In particular, consider the situation where a parameter $\s \in [\thetabar]$ is only payoff equivalent to $\sran$ for the fixed point strategy $\qbar$ but not for strategies in a local neighborhood of $\qbar$. In this case, local consistency is not satisfied, and payoffs generated by any local perturbation of the strategy will allow players to distinguish $\s$ from $\sran$ with positive probability. Consequently, beliefs and strategies may leave the local neighborhood of $(\thetabar, \qbar)$ with positive probability when it is locally perturbed. 

Theorem \ref{theorem:stability} and Proposition \ref{prop:complete_local} show that complete information fixed points are robust to such local perturbations, but other fixed points may not be. While these two results do not rule out the possibility that learning may still converge to fixed points that do not have complete information, these fixed points are non-robust unless they are locally stable. In next section, we further analyze conditions, under which learning dynamics converge to a complete information equilibrium.



\subsection{Complete Learning}\label{subsec:complete}
We say that learning is complete if the convergent fixed point strategy profile $\qbar$ is a complete information equilibrium (i.e. $\qbar = \qsran$). We now answer two questions: When is learning complete? If learning is not complete (i.e. $\qbar \neq \qsran$), can we still find the complete information equilibrium? 

Firstly, if the convergent fixed point belief assigns probability 1 to a single parameter (i.e. the support set $[\thetabar]$ is a singleton set), then the unique parameter in $[\thetabar]$ must be the true parameter $\sran$. In this case, we know that the fixed point must be a complete information fixed point $(\thetasran, \qsran)$, and thus learning is complete. Besides, from Proposition \ref{prop:global}, we know that learning is guaranteed to converge to a complete information fixed point if and only if the true parameter is identifiable at equilibrium.

 Secondly, if the convergent fixed point belief assigns positive probability to more than one parameters (i.e. the support set $[\thetabar]$ is not a singleton set), then the fixed point belief does not have complete information of the true parameter. In this case, Proposition \ref{prop:equivalent_complete} below provides a sufficient condition on the fixed point belief and players' payoff functions to ensure that the fixed point strategy profile is a complete information equilibrium. Here, learning is still complete, although the belief does not identify the true parameter.

 
 

 \begin{proposition}\label{prop:equivalent_complete}
For a fixed point $\(\thetabar, \qbar\)$, $\qbar = \qsran$ if 
\begin{itemize}
    \item[(i)] \emph{Local consistency.} The fixed point belief $\thetabar$ forms a consistent payoff estimate in a local neighborhood of $\qbar$, i.e. $\exists \xi >0$ such that $[\thetabar] \subseteq \Sequiv(\q)$ for any $\q \in N_{\xi}(\qbar)$;
    \item[(ii)] \emph{Payoff concavity.} The payoff function $\usi(\qi, \qmi)$ is concave in $\qi$ for all $\qmi \in \Qmi$, all $\i \in \I$ and all $\s \in [\thetabar]$. 
\end{itemize}
\end{proposition}


For any fixed point $\(\thetabar, \qbar\)$, since $\qbar_i$ is a best response strategy of $\qbar_{\mi}$, $\qbar_i$ is a local maximizer of the expected payoff function $\mathbb{E}_{\thetabar}[u_i^s(\qi, \qbar_{\mi})]$. The local consistency condition \emph{(i)} in Proposition \ref{prop:equivalent_complete} ensures that the value of the expected payoff function is identical to the one corresponding to the true parameter $\sran$ for any $\qi$ belonging to a small neighborhood of $\qbar_i$. Therefore, $\qbar_i$ must be a local maximizer of the payoff function with the true parameter $u_i^{\sran}(\qi, \qbar_{\mi})$. Condition \emph{(ii)} further provides that payoffs are concave functions of $\q_i$, and thus $\qbar_i$ must also be a global maximizer of $u_i^{\sran}(\qi, \qbar_{\mi})$. Therefore, \emph{(i)} and \emph{(ii)} together ensure that the fixed point strategy $\qbar$ is a complete information equilibrium, although $\thetabar$ may not fully identify the true parameter $\sran$.\footnote{
We note that for continuous and differentiable utility functions, the local consistency condition is equivalent to requiring that the gradient of the expected utility function based on the fixed point belief is identical to the gradient of the true utility function in a local neighborhood of the fixed point, i.e. $\mathbb{E}_{\thetabar}[\frac{\partial u_i^s(q)}{\partial q_i}] =\frac{\partial u_i^{\sran}(q)}{\partial q_i}$ for all $q \in N_{\xi}(\qbar)$. In classical no-regret learning dynamics (e.g. \citet{mertikopoulos2019learning}), the convergence of strategies requires that the payoffs are strictly concave, and the estimate of the gradient of the utility function is unbiased in each stage. In our belief-based no-regret learning \eqref{eq:update_belief} -- \eqref{eq:gradient}, the estimated gradient computed by the belief in each stage may not be consistent with the gradient of the true utility function. From Proposition \ref{prop:equivalent_complete}, we can conclude that in strictly concave game, when players choose belief-based no-regret learning, strategies converge to a complete information equilibrium if the gradient estimate is consistent in a local neighborhood of the fixed point. }


To summarize, we can guarantee that learning is complete in two cases: either the convergent belief identifies a single parameter, or conditions \emph{(i)} and \emph{(ii)} in Proposition \ref{prop:equivalent_complete} are satisfied. When neither of these cases apply to the learning dynamics, we can conclude that the convergent $(\thetabar, \qbar)$ is definitely not a complete information fixed point, but $\qbar$ may or may not be a complete information equilibrium.


    
    Typically, literature on continuous games assumes that the payoff concavity condition is satisfied.\footnote{In continuous games, payoff concavity guarantees the existence of pure equilibrium, and thus is typically assumed (\citet{rosen1965existence}).} In these games, if $\thetabar$ is not locally consistent, then there must exist at least one parameter $\s \in [\thetabar]$ such that $s$ is not payoff equivalent to $\sran$ in any local neighborhood of $\qbar$. Then, \emph{local exploration}, that is randomly perturbing strategies in a local neighborhood of $\qbar$, can distinguish such $\s$, and thus leads learning to a new fixed point $(\thetabar', \qbar')$ where $\thetabar'$ is locally consistent. Since both conditions in Proposition \ref{prop:equivalent_complete} are satisfied for the new fixed point, we can conclude that $\qbar'$ is a complete information Nash equilibrium. Therefore, in games with concave payoff functions, local exploration can find a complete information equilibrium. 
    
More generally, when payoff concavity is not satisfied, $\qbar'$ may not be a complete information equilibrium, and one needs a different approach to identify the true parameter in the support set $[\thetabar']$. Unfortunately, local exploration no longer helps since all parameters in $[\thetabar']$ are payoff equivalent with $\sran$ in local neighborhood of $\qbar'$. Instead, one needs to distinguish each pair of $s, s' \in [\thetabar']$ by exploring strategies for which these two parameters are not payoff equivalent, and such strategies may not be in local neighborhood of $\qbar'$. 
    
    \color{black}
    
    


\subsection{Examples}\label{subsec:example}
We present three examples to further illustrate our results. 

\vspace{0.2cm}
\noindent\textbf{Example 1. (Cournot competition)} A set of $\I$ firms produce an identical product and compete in a market. In each stage $\t$, firm $\i$'s strategy is their production level $\qit\in [0, 3]$. The price of the product is $p^{\t}=\alpha^\s - \beta^s \(\sum_{\i \in \I} \qit\)+ \epsilon^s$, where $\s=\(\alpha^s, \beta^s\)$ is the unknown parameter vector in the price function, and $\epsilon$ is a random variable with zero mean. The set of parameter vectors is $\S=\{s_1, s_2\}$, where $\s_1 = \(2, 1\)$ and $\s_2= \(4, 3\)$. The true parameter is $\sran=s_1$. The marginal cost of each firm is 0. The payoff of firm $\i$ in stage $\t$ is $\ct_i=\qit \(\alpha^\s - \beta^\s \(\sum_{\i \in \I} \qit\)+ \epsilon^s\)$ for each $\s \in \S$.


The information platform does not need to observe the production level or the payoff of each firm. $i \in I$. Instead, the platform can update belief $\thetat$ based on the total production $\tilde{q}^k = \sum_{\i \in \I} \qit$, and the price $p^\t = \ct_i/\qt_i$ for all $i \in I$. This is due to the fact that $\phi^s(\ct|\qt) = \tilde{\phi}^s(p^\t|\tilde{\q}^\t)$ for all $\s \in \S$, where $\tilde{\phi}^s(p^\t|\tilde{\q}^\t)$ is the probability density function of the price given the total production. The belief update given $(\tilde{q}^k, p^k)_{k=1}^{\infty}$ is as follows: \begin{align*}
    \theta^{k_{t+1}}(\s)&= \frac{\theta^{\kt}(\s) \prod_{\t=\kt}^{\t_{t+1}-1}\tilde{\phi}^s(p^\t|\tilde{\q}^\t)}{\sum_{s' \in \S} \theta^{\kt}(\s') \prod_{\t=\kt}^{\t_{t+1}-1} \tilde{\phi}^{s'}(p^\t|\tilde{\q}^\t)}, \quad \forall \s \in \S, \quad \forall t=1, 2, \dots. 
\end{align*}

In this game, the expected payoff function $\mathbb{E}_{\theta}[u_i^s(q_i, q_{-i})] = \sum_{\s \in \S} \theta(s) \qi \(\alpha^\s - \beta^\s \(\sum_{\i \in \I} \qi\)\)$ is continuous in $\theta$ and concave in $q_i$ for all $i \in I$. Thus, the best response correspondence $BR_i(\theta, \qmi)$ is upper-hemicontinuous in $\theta$ and $\qmi$ for each $i \in I$ following from the Berge's maximum theorem. As a result, the strategy updates \eqref{eq:br} and \eqref{eq:br_diminish} satisfy Assumption \ref{as:uh}. Additionally, since the utility function $\mathbb{E}_{\theta}[u_i^s(q_i, q_{-i})] $ is continuously differentiable with respect to the strategy profile $q$, the update \eqref{eq:gradient} is also upper-hemicontinuous in $\theta$ and $q$ for all $i \in I$, and thus satisfies Assumption \ref{as:uh}. Furthermore, for any $\theta$, the Cournot game $G(\theta)$ has a concave potential function given by: 
\begin{align*}
    \Psi_{\theta}(q) = \mathbb{E}_{\theta}[\alpha^s] \left(\sum_{i \in I} q_i \right) - \mathbb{E}_{\theta}[\beta^s] \left(\sum_{i \in I} q_i^2 \right) - \mathbb{E}_{\theta}[\beta^s] \left(\sum_{1 \leq i \leq j \leq |I|} q_i q_j\right). 
\end{align*}
Therefore, the strategy updates -- \eqref{eq:br}, \eqref{eq:br_diminish}, and \eqref{eq:gradient} with respect to a static belief $\theta$ converge 
 to a Nash equilibrium in $\EQ(\theta)$, i.e.  \eqref{eq:br}, \eqref{eq:br_diminish}, and \eqref{eq:gradient} satisfies Assumption \ref{as:convergence} (Table \ref{tab:summary}). \textcolor{blue}{Additionally, since for any $\theta$, the equilibrium set 
 $\EQ(\theta)$ is unique, Assumption \ref{as:isolation} is satisfied.}

From Theorem \ref{theorem:convergence}, the states of the learning dynamics with strategy updates \eqref{eq:br}, \eqref{eq:br_diminish}, and \eqref{eq:gradient} converge to a fixed point with probability 1. The complete information fixed point is $\thetasran=\(1,0\)$ and $\qsran=\(2/3, 2/3\)$. Additionally, $\theta^{\dagger}= \(0.5, 0.5\)$ and $\q^{\dagger}=\(0.5, 0.5\) \in \EQ(\theta^{\dagger})$ is also a fixed point since $[\theta^{\dagger}] = \Sequiv(\q^{\dagger})=\{\s_1, \s_2\}$. Note that at $\q^{\dagger}$, the parameter $s_2$ leads to identical price distribution as $s_1$, and thus is payoff equivalent. In fact, since any $\theta \neq \thetasran$ must include $\s_2$ in the support set, one can show that $\q^{\dagger}= \(0.5, 0.5\)$ is the only strategy profile for which $\s_1$ and $\s_2$ are payoff-equivalent. Thus, there does not exist any other fixed points apart from $\(\thetasran, \qsran\)$ and $\(\theta^{\dagger}, \q^{\dagger}\)$; i.e. $\FP=\left\{(\thetasran, \qsran), (\theta^{\dagger}, \q^{\dagger})\right\}$.
 
Since the complete information fixed point is not the unique fixed point, no fixed point is globally stable (Proposition \ref{prop:global}). We know from Proposition \ref{prop:complete_local} that the complete information fixed point $\thetasran=(1, 0)$, $\qsran=\(2/3, 2/3\)$ is locally stable. On the other hand, the other fixed point $\theta^{\dagger}=\(0.5, 0.5\)$ and $\q^{\dagger}=\(0.5, 0.5\)$ does not satisfy the local consistency condition since the two parameters $\s_1$ and $\s_2$ can be distinguished when the strategy is perturbed in any local neighborhood of $\q^{\dagger}$. Since the utility functions are concave, local exploration around $\q^{\dagger}$ can identify the true parameter, and leads to the complete information equilibrium. 

\vspace{0.2cm}

\noindent\textbf{Example 2.(Zero sum game)}
  Two players $\i  \in \{1, 2\}$ repeatedly play a zero-sum game with identical convex and closed strategy sets $\Q_1=\Q_2=[0,6]$. For any strategy profile $\q$, the payoff of each player is $\c_1 = - \c_2 = v^\s(\q) + \epsilon^\s$, where 
  \begin{align*}
    v^\s(\q)= \(\max\(|\qt_1 - \qt_2|, \s\) - \s\)^2 - 2(\qt_1)^2\textcolor{blue}{+ \frac{1}{2}(q_2^k - 2)^2},
\end{align*}
and $\s \in \S = \{1, 3, 5\}$ is the unknown parameter. The true parameter $\sran=3$. Belief is updated by an information platform based on the strategy profiles $(\qt)_{\t=1}^{\infty}$ and the realized payoffs $(y_1^k, y_2^k)_{k=1}^{\infty}$ as in \eqref{eq:update_belief}. 

For any $\theta$, the game $G(\theta)$ is a zero-sum game with expected value function $\mathbb{E}_{\theta}[v^s(q)]$ that is continuous in $\theta$, concave in $q_1$ and convex in $q_2$. Thus, the strategy updates \eqref{eq:br_diminish} and \eqref{eq:gradient} satisfy Assumptions \ref{as:uh} and \ref{as:convergence} (Table \ref{tab:summary}). Moreover, for any $\theta$ such that $\theta(1)=0$, the equilibrium is unique $(0, 2)$. For any $\theta$ such that $\theta(1)>0$, the equilibrium strategy profile is $(0, (2+2\theta(1))/(2\theta(1)+1))$. Therefore, Assumption \ref{as:isolation} is satisfied. 

From Theorem \ref{theorem:convergence}, the sequence of states converges to a fixed point w.p. 1. The set of complete information fixed points is $\thetasran=\(0, 1, 0\)$ and $\EQ(\thetasran)= \{(0,2)\}$. Apart from the complete information fixed points, any $\theta^{\dagger} \in \{\Delta(\S) | \theta^{\dagger}(1)=0\}$ and $\q^{\dagger} =(0,2)$ is also a fixed point. This is because for any belief $\theta^{\dagger}$ that assigns zero probability on $\s=1$, $\q^{\dagger}_1=0$ and $\q^{\dagger}_2=2$ is an equilibrium, and the two parameters $\s=3$ and $\s=5$ are payoff equivalent at $\q^{\dagger}$. 

We can check that conditions (i) and (ii) in Proposition \ref{prop:equivalent_complete} are satisfied by any fixed point $\(\theta^{\dagger}, \q^{\dagger}\)$. Thus, any fixed point strategy is a complete information equilibrium although $\theta^{\dagger}$ is not a complete information belief. Therefore, learning is guaranteed to be complete in this game. 

Since the complete information fixed point is not unique, no fixed point is globally stable. Moreover, by setting $\epsilon=1/2$ and $\delta= 6$, we can check that all fixed points in $\FP$ satisfy the two conditions in Assumption \ref{as:stability}, and thus are locally stable (Theorem \ref{theorem:stability}).  

\color{black}
\vspace{0.2cm}
\noindent\textbf{Example 3. (Investment game)}
Two players repeatedly play an investment game. In each stage $\t$, the strategy $\qit \in [0, 1]$ is the non-negative level of investment of player $\i$. Given the strategy profile $\qt = \(\qt_1, \qt_2\)$, the return of a unit investment is randomly realized according to $r^\t=\s +\qt_1+ \qt_2 + \epsilon^\s$, where $\s \in \S=\{0,1,2\}$ is the unknown parameter that represents the average baseline return and $\epsilon^s$ is the noise term. The true parameter is $\sran=1$. The stage cost of investment for each player is $3 \(\qit\)^2$. Therefore, the payoff of each player $\i \in \I$ is $\ct_i= \qit(\s +\qt_1+ \qt_2+\epsilon^\s) - 3 \(\qit\)^2 = \qit(\s -2 \qit+ \qt_{\mi} + \epsilon^\s)$ for all $\s \in \S$. Following similar argument as in Example 1, the information platform does not need to observe the strategy of each player and their realized payoffs. Instead, the platform can update the belief $\thetat$ based on the total investment $\tilde{q}^k = \qt_1+\qt_2$ and the realized unit investment return $r^\t$ in each stage $\t$. 

For any $\theta$, the expected utility function $\mathbb{E}_{\theta}[u_i^s(q)]$ is continuous in $\theta$ and $q$. Moreover, the game $G(\theta)$ is a supermodular game (i.e. $\partial^2 \mathbb{E}_{\theta}[u_i^s(q)]/\partial q_i \partial q_j = 1 >0$), and it is also dominance solvable with unique equilibrium strategy profile $\EQ(\theta) = \{(\mathbb{E}_{\theta}[s]/3, \mathbb{E}_{\theta}[s]/3)\}$. Therefore, all three best response dynamics \eqref{eq:sbr} -- \eqref{eq:br_diminish} satisfy Assumptions \ref{as:uh} and \ref{as:convergence} (Table \ref{tab:summary}). \textcolor{blue}{Since the equilibrium set is a singleton set for all $\theta$, Assumption \ref{as:isolation} is also satisfied.} Thus, states converge to a fixed point with probability 1. In this game, since $\Sequiv(\q)=\{\sran=1\}$ for any $\q \in \Q$, the unique fixed point is the complete information fixed point, i.e. $\FP = \{\(\thetasran, \qsran\)= \(\(0, 1, 0\), \(1/3, 1/3\)\)\}$. From Proposition \ref{prop:global}, we know that this complete information fixed point is globally stable. 


\section{Proofs of Main Results}\label{sec:proof}
In this section, we present the proofs of our convergence and local stability results (Theorems \ref{theorem:convergence} and \ref{theorem:stability}). Recall that in \eqref{eq:update_belief}, the beliefs are updated at a random subsequence of stages $\(\kt\)_{k=1}^{\infty}$. For the ease of presentation in our proof, we introduce an auxiliary belief sequence $\(\thetatilt\)_{\t=1}^{\infty}$ that is updated in every stage:
\begin{align}\label{eq:update_belief_always}
    \tilde{\theta}^1= \thetazero,\text{ and } ~\tilde{\theta}^{\t+1}(\s) = \frac{\thetatilt(\s)\phi^s(\ct|\qt)}{\sum_{\s' \in \S}\thetatilt(\s')\phi^{s'}(\ct|\qt)}, \quad \forall \s \in \S, \quad \forall \t=1, 2, \dots
\end{align}
From \eqref{eq:update_belief}, we know that
\begin{align}\label{eq:relationship}
    \thetat=\left\{\begin{array}{ll}
    \thetatilt, &\quad  \text{if  }\t=\kt, \quad \forall t=1, 2, \dots, \\
    \theta^{\t-1}, & \quad \text{otherwise}.
    \end{array}
    \right.
\end{align}
We can also write the auxiliary belief ratio as follows: 
\begin{align}\label{eq:ratio}
    \frac{\thetatilt(s)}{\thetatilt(\sran)} = \frac{\tilde{\theta}^1(s)}{\tilde{\theta}^1(\sran)} \cdot \prod_{j=1}^{k-1} \frac{\phi^s(\cj|q^j)}{\phi^{\sran}(\cj|q^j)}
\end{align}Since the payoffs and strategies are public information, we show in the following lemma that the auxiliary belief ratio is a martingale sequence. The proof of Lemma \ref{lemma:thetatil} is included in Appendix \ref{apx:proof}.
\begin{lemma}\label{lemma:thetatil}
For any $s \in S$, the sequence of belief ratios $\left(\frac{\thetatilt(s)}{\thetatilt(\sran)}\right)_{\t=1}^{\infty}$ is a martingale sequence. 
\end{lemma}
Our proofs of the convergence and stability results build on the martingale property of belief ratio. In particular, we utilize the martingale property to show the convergence of beliefs in Theorem \ref{theorem:convergence}. In our proof of local stability in Theorem \ref{theorem:stability}, we rely on the martingale property to compute the upper bound of the probability that beliefs leave a local $\bar{\epsilon}$-neighborhood of $\thetabar$, and derive conditions on the initial belief to ensure that this upper bound does not exceed $\gamma$ as in the definition of local stability (Definition \ref{def:local}).

\subsection{Proof of convergence}

We prove Theorem \ref{theorem:convergence} in three steps: Firstly, we prove that the sequence of beliefs $\(\thetat\)_{\t=1}^{\infty}$ converges to a fixed point belief $\thetabar \in \Delta\(\S\)$ w.p. 1 by applying the martingale convergence theorem (Lemma \ref{lemma:theta}). Secondly, we show that under Assumptions \ref{as:uh} and \ref{as:convergence}, the strategies $\(\qt\)_{\t=1}^{\infty}$ in our learning dynamics also converge. This convergent strategy is an equilibrium corresponding to the fixed point belief $\thetabar$ (Lemma \ref{lemma:q}). Finally, we prove that the belief of any $\s \in \S$ that is not payoff-equivalent to $\sran$ given $\qbar$ must converge to $0$ with rate of convergence governed by \eqref{eq:rate} (Lemma \ref{lemma:consistency}). Hence, we can conclude that beliefs and strategies induced by the learning dynamics converge to a fixed point $\(\thetabar, \qbar\)$ that satisfies \eqref{eq:fixed_point_def} w.p. 1.

\begin{lemma}\label{lemma:theta}
$\lim_{\t \to \infty} \thetat=\thetabar$ w.p. 1, where $\thetabar  \in\Delta(S)$.
\end{lemma}

Lemma \ref{lemma:theta} directly follows from Lemma \ref{lemma:thetatil}. We include the proof in Appendix \ref{apx:proof}.

We next prove the convergence of strategies. For every $K=1, 2, \dots $,  we construct an auxiliary strategy sequences $\(\qhatt_K\)_{j=1}^{\infty}$ such that $\qhatt_K$ are identical to $\q^j$ in the original sequence generate by the learning dynamics up to the stage $\K$ (i.e. $\qhatt_K=\q^j$ for all $j=1, \dots, \K$), and the remaining strategies $\(\qhatt_K\)_{j=\K+1}^{\infty}$ are induced  by updates associated with the fixed point belief $\thetabar$ (instead of the repeatedly updated belief sequence $\(\thetat\)_{\t=\K+1}^{\infty}$). In particular, for each $j \geq K$, $\hat{q}^{j+1}_K$ is obtained as follows: 
\begin{align}\label{eq:auxiliary}
\qhat^{j+1}_K = \argmin_{\q\in F(\thetabar, \qhatt_K)} \|\q- \tilde{q}^{j+1}\|, \quad \text{such that} \quad  \tilde{\q}^{j+1} = \argmin_{\tilde{\q}\in F(\thetabar, \q^{j})} \|\tilde{\q} - \q^{j+1}\|.
\end{align}
From \eqref{eq:auxiliary}, $\hat{q}^{j+1}_K$ is constructed in two steps: Firstly, given the original strategy $q^j$ and the fixed point belief $\thetabar$, we compute $\tilde{\q}^{j+1}$ as the the strategy updated from $q^j$ with respect to $\thetabar$ (i.e. $\tilde{\q}^{j+1} \in F(\thetabar, \q^{j})$) that is the closest to the original strategy $\q^{j+1}$. Secondly, $\qhat^{j+1}_K$ is updated from the auxiliary strategy $\qhat^{j}_K$ in stage $j$ with the fixed point belief $\thetabar$ (i.e. $\qhat_K^{j+1} \in F(\thetabar, \qhat_K^{j})$), and $\qhat^{j+1}_K$ is the closest to $\tilde{q}^{j+1}$ among all strategies in the set $F(\thetabar, \qhat^{j}_K)$. 

The two-step construction ensures that $\(\qhatt_K\)_{j=\K+1}^{\infty}$ has the smallest Euclidean distance with the original strategy sequence among all strategy sequences that are induced by updates with $\thetabar$. In fact, we can show that, as the beliefs converge to $\thetabar$ (Lemma \ref{lemma:theta}), the distance between the auxiliary strategy sequence and the original strategy sequence also converges to zero as $K \to \infty$ under Assumption \ref{as:uh}. Additionally, under Assumption \ref{as:convergence}, the auxiliary strategy sequence must converge to an equilibrium $\qbar \in \EQ(\thetabar)$, and thus the original sequence $\(\qt\)_{\t=1}^{\infty}$ also converges to $\qbar$.

\begin{lemma}\label{lemma:q}
The sequence of strategies $(q^k)_{k=1}^{\infty}$ converges with probability 1. Moreover, the convergent strategy profile $\qbar \in \EQ(\thetabar)$. 
\end{lemma}

\medskip

\noindent\emph{Proof of Lemma \ref{lemma:q}.} 
Consider any sequence of states $(\theta^j, \q^j)_{j=1}^{\infty}$ induced by the learning dynamics \eqref{eq:update_belief} -- \eqref{eq:generic}. For any $K$, we construct the auxiliary strategy sequence $(\qhat^{j}_K)_{j=1}^{\infty}$ from \eqref{eq:auxiliary}. For any $j > \K$, we have: 
\begin{align}\label{eq:dist}
    \|\q^{j+1}-\tilde{\q}^{j+1}\|= \dist\(\q^{j+1}, F(\thetabar, q^j)\),~ \|\tilde{\q}^{j+1} -\qhat^{j+1}_K\|= \dist\(\tilde{q}^{j+1}, F(\thetabar, \qhatt_K)\).
\end{align}
We next show by mathematical induction that for any $\ell\geq 1$, $\lim_{\K \to \infty} \|\q^{\K+\ell}- \qhat^{\K+\ell}_K\| =0$. To begin with, for $\ell=1$, we have
\begin{align}
    &\|\q^{\K+1}- \qhat^{\K+1}_K\| \leq \|\q^{\K+1}- \tilde{q}^{\K+1}\|+ \|\tilde{\q}^{\K+1} - \qhat^{\K+1}_K\| \notag \\
    \stackrel{\eqref{eq:dist}}{=}&\dist\(\q^{K+1}, F\(\thetabar, \q^{\K}\)\) + \dist\(\tilde{q}^{\K+1}, F\(\thetabar, \qhat^{\K}_K\)\). \label{eq:Kl}
\end{align}
Since $\thetat$ converges to $\thetabar$ (Lemma \ref{lemma:theta}), $F(\theta, \q)$ is upper hemicontinuous in $\theta$ (Assumption \ref{as:uh}), and $\q^{K+1} \in F(\theta^{K+1}, \q^K)$, we know that $\lim_{\K \to \infty}\dist\(\q^{K+1}, F\(\thetabar, \q^{\K}\)\)=0$. Additionally, since $\qhat^K_K=\q^K$ and $\tilde{\q}^{K+1} \in F(\thetabar, \q^{\K})$, $\dist\(\tilde{q}^{\K+1}, F\(\thetabar, \qhat^{\K}_K\)\)=0$. Therefore, $\lim_{\K \to \infty} \|\q^{\K+1}- \qhat^{\K+1}\| =0$.

Now, assume that $\lim_{\K \to \infty} \|\q^{\K+\ell}- \qhat_K^{\K+\ell}\| =0$ for some $\ell\geq 1$, we need to show that $\lim_{\K \to \infty} \|\q^{\K+\ell+1}- \qhat_K^{\K+\ell+1}\| =0$. Similar to \eqref{eq:Kl}, we have
\begin{align*}
  \|\q^{\K+\ell+1}- \qhat_K^{\K+\ell+1}\| \leq \dist\(\q^{K+\ell+1}, F\(\thetabar, \q^{\K+\ell}\)\) + \dist\(\tilde{q}^{\K+\ell+1}, F\(\thetabar, \qhat_K^{\K+\ell}\)\).
\end{align*}
Analogous to $\ell=1$, since $F(\theta, \q)$ is upper hemicontinuous in $\theta$, $\lim_{\K \to \infty} \dist\(\q^{K+\ell+1}, F\(\thetabar, \q^{\K+\ell}\)\) =0$. Additionally, since $\lim_{\K \to \infty} \|\q^{\K+\ell}- \qhat^{\K+\ell}_K\| =0$, $F(\theta, \q)$ is upper hemicontinuous in $\q$ (Assumption \ref{as:uh}), and $\tilde{q}^{\K+\ell+1} \in F\(\thetabar, \q^{\K+\ell}\)$, we know that $\lim_{\K \to \infty}  \dist\(\tilde{q}^{\K+\ell+1}, F\(\thetabar, \qhat_K^{\K+\ell}\)\)=0$.  Therefore, we have $\lim_{\K \to \infty} \|\q^{\K+\ell+1}- \qhat^{\K+\ell+1}_K\| =0$. By mathematical induction, we conclude that for any $\ell\geq 1$, $\lim_{\K \to \infty} \|\q^{\K+\ell}- \qhat^{\K+\ell}_K\|=0$. 

Under Assumption \ref{as:convergence}, we know that given any initial strategy $\hat{q}^1 \in Q$, the sequence of strategies $(\hat{q}^k)_{k=1}^{\infty}$ induced by updates \eqref{eq:generic} with static belief $\thetabar$ converges to the equilibrium set $\EQ(\thetabar)$. Then, for any $\epsilon>0$ and any initial strategy $\hat{q}^1 \in Q$, there exists a finite integer $L(\hat{q}^1)$ such that $D(\hat{q}^k, \EQ(\thetabar)) < \epsilon/2$ for any $k \geq L(\hat{q}^1)$. Since the strategy set $Q$ is a closed and bounded set, we take $L = \max_{\hat{q}^1 \in \Q} L(\hat{q}^1)$, and we know that $L$ is finite.

Next, for any $k \geq L$, the original strategy sequence $(q^k)_{k=1}^{\infty}$ satisfies 
\begin{align*}
D(q^k, \EQ(\thetabar)) \leq \|q^k - \hat{q}_{k-L}^k\|+ D(\hat{q}_{k-L}^k, \EQ(\thetabar)),  
\end{align*}
where $\hat{q}_{k-L}^k$ is the $k$-th strategy in the auxiliary sequence $(\hat{q}_{k-L}^{j})_{j=1}^{\infty}$. Recall that the auxiliary strategy sequence $(\hat{q}_{k-L}^{j})_{j=1}^{\infty}$ is such that $\hat{q}_{k-L}^{j}=q^j$ for $j =1, \dots, k-L$, and $\hat{q}_{k-L}^{j}$ is constructed with respect to the fixed point belief $\thetabar$ as in \eqref{eq:auxiliary} for $j=k-L+1, \dots, $. Thus, for any $k \geq L$, $\hat{q}_{k-L}^k$ represents a strategy profile resulting from $L$ strategy updates with respect to $\thetabar$ starting from the initial profile $q^{k-L}$. Then, $D(\hat{q}_{k-L}^k, \EQ(\thetabar)) < \epsilon/2$ for all $k \geq L$. Moreover, since $\lim_{k \to \infty} \|\q^{k+\ell}- \qhat^{k+\ell}_k\|=0$ for any $\ell$, we know that $\lim_{k \to \infty} \|q^k - \hat{q}_{k-L}^k\|=0$. Hence, there must exist $K$ such that for any $k \geq K+L$, $\|q^k - \hat{q}_{k-L}^k\| < \epsilon/2$. Consequently, we have that for any $k \geq K+L$, $D(q^k, \EQ(\thetabar)) \leq \|q^k - \hat{q}_{k-L}^k\|+ D(\hat{q}_{k-L}^k, \EQ(\thetabar)) < \epsilon$, i.e. $\lim_{k \to \infty} D(q^k, \EQ(\thetabar))=0$. The sequence of strategies $\left(q^k\right)_{k=1}^{\infty}$ converges to the equilibrium set $\EQ(\thetabar)$. 

If $\EQ(\thetabar)$ is a singleton set, then we can conclude that $(\qt)_{k=1}^{\infty}$ converges. If $\EQ(\thetabar)$ contains multiple equilibria, it remains to prove the convergence of the strategy sequence $(\qt)_{k=1}^{\infty}$ to one of the equilibria. Assume for the sake of contradiction that $(\qt)_{k=1}^{\infty}$ does not converge. Since $\lim_{k \to \infty} D(q^k, \EQ(\thetabar))=0$, there must exist $K_a$ such that for any $k> K_a$, we can find a $q^{k^\dagger} \in \EQ(\thetabar)$ such that $\|q^k - q^{k\dagger}\| < \delta/4$, where $\delta$ is the lower bound of the distance between any pair of equilibrium as in Assumption \ref{as:isolation}. We note that $q^{k\dagger}$ can be viewed as the equilibrium strategy in the set $\EQ(\thetabar)$ that is the closest to $\qt$, and $q^{k\dagger}$ may be different for different $k> K_a$. In fact, under the assumption that the strategy sequence does not converge, we know that for any $K$, there must exist $k, k+1 > \max\{K, K_a\}$ such that $\|q^k - q^{k\dagger}\| 
<\delta/4$ and $\|q^{k+1} - q^{k+1\dagger}\| < \delta/4$, where $q^{k\dagger} \neq q^{k+1\dagger}$. Otherwise, the strategy sequence $(q^k)_{k=1}^{\infty}$ will remain in the local neighborhood of one equilibrium after a finite stage, which implies the convergence to that equilibrium given that $\lim_{k \to \infty} D(q^k, \EQ(\thetabar))=0$ and each equilibrium is isolated. Moreover, for any such pair of stages $k, k+1$, since $\|q^{k\dagger} -q^{k+1\dagger}\| \geq \delta$, we must have $\|q^{k+1} - q^{k\dagger}\| \geq |\|q^{k+1}- q^{k+1 \dagger}\| - \|q^{k\dagger} - q^{k+1 \dagger}\| | \geq 3\delta/4$.

Given Assumptions \ref{as:uh} and \ref{as:isolation}, we argue that for any $q^{\dagger} \in \EQ(\thetabar)$, there must exist $\xi_{a}, \xi_{b}>0$ such that for any $q$ that satisfies $\|q- q^{\dagger}\|< \xi_a$ and any $\theta$ that satisfies $\|\theta - \thetabar\|< \xi_b$, the updated strategy $q' = F(\theta, q)$ satisfies $\|q' - q^{\dagger}\|< \delta/4$. This is due to the fact that (i) $q^{\dagger} = F(\thetabar, q^{\dagger})$ ($q^{\dagger}$ is a fixed point of the strategy update with respect to $\thetabar$, and it is an isolated equilibrium in $\EQ(\thetabar)$) (Assumption \ref{as:isolation}), and (ii) $F(\cdot, \cdot)$ is upper hemicontinuous in both $\theta$ and $\q$ (Assumption \ref{as:uh}) and unique in the local neighborhood of $(\thetabar, q^{\dagger})$ (Assumption \ref{as:isolation}).

Given that $\thetat$ converges to $\thetabar$ with probability 1 (Lemma \ref{lemma:theta}) and $\lim_{k \to \infty} D(q^k, \EQ(\thetabar))=0$, we know that there exists $K_b>0$ such that for any $k >K_b$, $q^k$ is within $\varepsilon = \min\{\xi_a, \delta/4\}$ distance to the closest equilibrium strategy $q^{k\dagger} \in \EQ(\thetabar)$, and the belief $\thetat$ is within $\xi_b$ distance to $\thetabar$. This implies that for any $k, k+1 > \max\{K, K_a, K_b\}$,  $\|q^{k} - q^{k\dagger}\|< \delta/4$ and $\|q^{k+1} - q^{k\dagger}\|< \delta/4$, which is a contradiction to the existence of $k, k+1$ where $\|q^k - q^{k\dagger}\| < \delta/4$  and $\|q^{k+1} - q^{k\dagger}\| \geq 3\delta/4$. Therefore, we have arrived at a contradiction, and we conclude that the sequence of strategies $(q^k)_{k=1}^{\infty}$ must converge. \QEDA

\vspace{0.2cm}

\color{black}


\begin{lemma}\label{lemma:consistency}
Any fix point $\(\thetabar, \qbar\)$ satisfies \eqref{eq:exclude_distinguished}. Furthermore, for any $\s \in \S\setminus \Sequiv(\qbar)$, if $\phibar^{\sran}(\chat|\qbar) \ll \phibar^{\s}(\chat|\qbar)$, then $\thetat(\s)$ satisfies \eqref{eq:rate}.
Otherwise, there exists a finite positive integer $\Tstar$ such that $\thetat(\s)=0$ for all $\t>\Tstar$ w.p. 1. 
\end{lemma}

Lemma \ref{lemma:consistency} is based on Lemmas \ref{lemma:theta} and \ref{lemma:q}. Although the realized payoffs $\(\ct\)_{\t=1}^{\infty}$ are not independently and identically distributed (i.i.d.) due to players' strategy updates, we can show that since $\qt$ converge to $\qbar$ (Lemma \ref{lemma:q}), as $k \to \infty$, the distribution of $\ct$ converges to an i.i.d. process with $\phi^{\s}(\ct|\qbar)$, which is the payoff distribution given the fixed point strategy $\qbar$, for each parameter $\s$ as $\t\to \infty$. This allows us to show that the belief eventually excludes parameters that are not in $\Sequiv(\qbar)$, and also establish the rate of convergence.


\medskip 
\noindent\emph{Proof of Lemma \ref{lemma:consistency}.} By iteratively applying the belief update in \eqref{eq:update_belief_always}, we can write:
\begin{align}\label{eq:sequence_update}
    \thetatilt(\s)= \frac{\thetatil^1(\s) \prod_{\j=1}^{\t-1}\phibar^{s}(\cj|\qj)}{\sum_{\s' \in \S} \thetatil^1(\s') \prod_{\j=1}^{\t-1}\phibar^{s'}(\cj|\qj)}, \quad \forall \s \in \S.
\end{align}
We define $\Phi^\s(\Costhis|\Loadhis)$ as the probability density function of the history of the realized payoffs $\Costhis=\(\c^j\)_{j=1}^{\t-1}$ conditioned on the history of strategies $\Loadhis= \(\q^j\)_{j=1}^{\t-1}$ prior to stage $\t$, i.e. $\Phi^\s(\Costhis|\Loadhis) \deleq \prod_{\j=1}^{\t-1}\phibar^{s}(\cj|\q^j)$. For any $s \in S\setminus \{\sran\}$, we can rewrite \eqref{eq:sequence_update} as follows: 
\begin{align}
    \thetatil^{\t}(\s)= \frac{\thetatil^1(\s) \Phi^\s(\Costhis|\Loadhis)}{\sum_{\s' \in \S} \thetatil^1(\s') \Phi^{\s'}(\Costhis|\Loadhis)} &\leq \frac{\thetatil^1(\s) \Phi^\s(\Costhis|\Loadhis)}{ \thetatil^1(\s) \Phi^{\s}(\Costhis|\Loadhis)+ \thetatil^1(\sran) \Phi^{\sran}(\Costhis|\Loadhis)}\notag\\
    &=\frac{\thetatil^1(\s) \frac{\Phi^\s(\Costhis|\Loadhis)}{\Phi^{\sran}(\Costhis|\Loadhis)}}{ \thetatil^1(\s) \frac{\Phi^{\s}(\Costhis|\Loadhis)}{\Phi^{\sran}(\Costhis|\Loadhis)}+ \thetatil^1(\sran)}.\label{eq:bound_theta}
\end{align}
For any $\s \in \S \setminus \Sequiv(\qbar)$, if we can show that the ratio $ \frac{\Phi^\s(\Costhis|\Loadhis)}{\Phi^{\sran}(\Costhis|\Loadhis)}$ converges to 0, then $\thetatil^{\t}(\s)$ must also converge to 0. We need to consider two cases:

\medskip
\noindent\emph{\underline{Case 1:}} $\phi^{\sran}(\cbar|\qbar) \ll \phi^{\s}(\cbar|\qbar)$. \\
In this case, the log-likelihood ratio can be written as: 
\begin{align}\label{eq:log-likelihood}
    \log \(\frac{\Phi^\s(\Costhis|\Loadhis)}{\Phi^{\sran}(\Costhis|\Loadhis)}\)=   \sum_{\j=1}^{\t-1} \log\(\frac{\phibar^{\s}(\cj|\qj)}{\phibar^{\sran}(\cj|\qj)}\).
\end{align}
For any $\s \in \S$, since $\phibar^{s}(\cj|\qj)$ is a continuous function of $\q^{j}$ for any realized $\cj$, $\etaj$ is also continuous in $\q^{j}$ for any $\cj$. In Lemma \ref{lemma:q}, we proved that $\(\qt\)_{\t=1}^{\infty}$ converges to $\qbar$. Then, $\etaj$ must converge to $\log\(\frac{\phibar^\s(\cj|\qbar)}{\phibar^{\sran}(\cj|\qbar)}\)$ as $j \to \infty$ for all realized payoff $\cj$. Therefore, from law of large numbers, we have: 
\begin{align*}
\lim_{\t \to \infty}\frac{1}{\t-1} \log \(\frac{\Phi^\s(\Costhis|\Loadhis)}{\Phi^{\sran}(\Costhis|\Loadhis)}\)&=\lim_{\t \to \infty}\frac{1}{\t-1} \sum_{j=1}^{\t-1} \etaj 
= \mathbb{E}\left[\etabar\right],~ w.p.~1.
\end{align*}

Note that for any $ \s \in \S \setminus \Sequiv(\qbar)$, the expectation of $\etabar$ can be written as: 
\begin{align*}
\mathbb{E}\left[\etabar\right]= \int_{\cbar}\phibar^{\sran}(\cbar|\qbar) \cdot  \log\(\frac{\phibar^\s(\cbar|\qbar)}{\phibar^{\sran}(\cbar|\qbar)}\)  d \cbar=-D_{KL}\(\phibar^{\sran}(\cbar|\qbar)||\phibar^\s(\cbar|\qbar)\)<0.
\end{align*}  

Then, for any $\s \in \S \setminus \Sequiv(\qbar)$, $\lim_{\t \to \infty} \frac{\Phi^\s(\Costhis|\Loadhis)}{\Phi^{\sran}(\Costhis|\Loadhis)}=0$. From \eqref{eq:bound_theta}, we know that $\lim_{\t \to \infty} \thetatil^{\t}(\s)=0$ for all $\s \in \S \setminus \Sequiv(\qbar)$. 
Finally, since $\thetatil^1(\s)>0$ for all $\s \in \S$, the true parameter $\sran$ is never excluded from the belief. For any $\s \in \S \setminus \Sequiv(\qbar)$, we have the following:
\begin{align*}
    &\lim_{\t \to \infty} \frac{1}{\t}\log\(\thetatilt(\s)\)=   \lim_{\t \to \infty} \frac{1}{\t} \log\(\thetatilt(\sran)\) +\lim_{\t \to \infty} \frac{1}{\t}\log\(\frac{\thetatilt(\s)}{\thetatilt(\sran)}\)\\
    &=\lim_{\t \to \infty} \frac{1}{\t}\log\(\frac{\thetatilt(\s)}{\thetatilt(\sran)}\) = \lim_{\t \to \infty} \frac{1}{\t}\log \(\frac{\thetatil^1(\s)}{\theta^1(\sran)}\)+\lim_{\t \to \infty} \frac{1}{\t} \log\(\frac{\Phi^\s(Y^{\t-1}|Q^{\t-1})}{\Phi^{\sran}(Y^{\t-1}|Q^{\t-1})}\)\\
    &= \mathbb{E} \left[\log\(\frac{\phibar^{\s}(\cbar|\qbar)}{\phibar^{\sran}(\cbar|\qbar)}\)\right]
    =-D_{KL}\(\phibar^{\sran}(\cbar|\qbar)||\phibar^\s(\cbar|\qbar)\), \quad w.p.~1. 
\end{align*}
From \eqref{eq:relationship}, we know that $\lim_{\t \to \infty} \frac{1}{\t}\log\(\thetat(\s)\) = -D_{KL}\(\phibar^{\sran}(\cbar|\qbar)||\phibar^\s(\cbar|\qbar)\)$. 

\medskip
\noindent\emph{\underline{Case 2:}} $\phi^{\sran}(\cbar|\qbar)$ is not absolutely continuous in $\phi^{\s}(\cbar|\qbar)$. \\
In this case, $\phi^\s(\cbar|\qbar)=0$ does not imply $\phi^{\sran}(\cbar|\qbar)=0$ with probability 1, i.e. $\pro\(\phi^\s(\cbar|\qbar)=0\)>0$, where $\pro\(\cdot\)$ is the probability of $\cbar$ with respect to the true distribution $\phi^{\sran}(\cbar|\qbar)$. Since the distributions $\phi^\s(\cbar|\q)$ and $\phi^{\sran}(\cbar|\q)$ are continuous in $\q$, the probability $\pro\(\phi^\s(\cbar|\q)=0\)$ must also be continuous in $\q$. Therefore, for any $\epsilon \in \(0, \pro\(\phi^\s(\cbar|\qbar)=0\)\)$, there exists $\delta>0$ such that $\pro\(\phi^\s(\c|\q)=0\) > \epsilon$ for all $\q \in \{\q|\|\q-\qbar\|<\delta\}$. 

From Lemma \ref{lemma:q}, we know that $\lim_{\t \to \infty} \qt = \qbar$. Hence, we can find a positive number $K_1>0$ such that for any $\t > K_1$, $\|\qt - \qbar\|<\delta$, and hence $\pro\(\phi^\s(\ct|\qt)=0\) > \epsilon$. We then have $\sum_{\t=1}^{\infty}\pro\(\phi^\s(\ct|\qt)=0\) = \infty$. Moreover, since the event $\{\phi^\s(\ct|\qt)=0\}$ is independent from the event $\{\phi^\s(\c^{\t'}|\q^{\t'})=0\}$ given any $\qt$ and $\q^{\t'}$, and any $\t, \t'$, we can conclude that $\pro\(\phi^\s(\ct|\qt)=0, \text{infinitely often}\)=1$ based on the second Borel-Cantelli lemma. Hence, $\pro\(\phi^\s(\ct|\qt)>0, ~\forall \t\)=0$. From the Bayesian update \eqref{eq:update_belief}, we know that if $\phi^\s(\ct|\qt)=0$ for some stage $\t$, then any belief of $\s$ updated after stage $\t$ is 0. Therefore, we can conclude that there exists a positive number $\Tstar > K_1$ with probability 1 such that $\thetat(\s)=0$ for any $\t > \Tstar$. 
    \QEDA

\subsection{Proof of local stability}

 From the local stability definition (Definition \ref{def:local}), for any $\bar{\epsilon}, \bar{\delta}>0$ and any probability $\gamma \in (0, 1)$, we need to characterize $\epsilon^1, \delta^1>0$ such that if the initial state is in $(\epsilon^1, \delta^1)$-neighborhood of the fixed point $(\thetabar, \EQ(\thetabar))$, then with probability higher than $\gamma$, the convergent state is in $(\bar{\epsilon}, \bar{\delta})$-neighborhood of $(\thetabar, \EQ(\thetabar))$.

Our proof of Theorem \ref{theorem:stability} starts with constructing an auxiliary $(\epsilonhat, \delta)$ neighborhood of $(\thetabar, \EQ(\thetabar))$ such that if all states remain in the auxiliary neighborhood, then the convergent state is guaranteed to be in the $(\bar{\epsilon}, \bar{\delta})$-neighborhood of $(\thetabar, \EQ(\thetabar))$ (Part \emph{(iii)} in Lemma \ref{lemma:constrained_set}). Here, $\epsilonhat \in (0, \epsilon)$ and $\epsilon$, and $\delta$ are chosen according to Assumption 2 to guarantee that the states in the auxiliary neighborhood satisfies the conditions of local consistency and local invariance (Part \emph{(i)} -- \emph{(ii)} in Lemma \ref{lemma:constrained_set}). 

\begin{lemma}\label{lemma:constrained_set}
Under Assumptions \ref{as:uh} -- \ref{as:stability},
\begin{enumerate}
    \item[(i)] For any $\loadepfin>0$, $\exists \epsilon' \in \(0, \epsilon\)$ such that any $\theta \in N_{\epsilon'}(\thetabar)$ satisfies $\EQ(\theta) \subseteq N_{\loadepfin}(\EQ(\thetabar))$.
    \item[(ii)] For any $\thetaepfin>0$, $F(\theta, \q) \subseteq N_{\delta}(\EQ(\thetabar))$ for all $\q \in N_{\delta}(\EQ(\thetabar))$ and all $\theta \in N_{\epsilonhat}\(\thetabar\)$, where $\epsilonhat = \min\{\epsilon, \epsilon', \thetaepfin\}$.  
    \item[(iii)] $\lim_{\t \to \infty} \pro\(\thetat \in \neighinftheta(\thetabar),\right.$ $ \left. \qt \in  \neighinfload(\EQ(\thetabar))\)\geq \pro\(\thetat \in  N_{\epsilonhat}(\thetabar), ~ \qt \in N_{\delta}(\EQ(\thetabar)), ~\forall \t\)$.
\end{enumerate}
\end{lemma}


We include the proof of Lemma \ref{lemma:constrained_set} in Appendix \ref{apx:proof}. Essentially,  \emph{(i)} follows from the fact that in continuous games, the equilibrium set $\EQ(\theta)$ is upper-hemicontinuous in $\theta$ (Lemma \ref{lemma:up} in Appendix \ref{apx:proof}). Then, we obtain \emph{(ii)} from Assumption \emph{(A3b)} that $N_{\delta}\(\EQ(\thetabar)\)$ is an invariant set of the strategy update for $\theta \in N_{\epsilon}(\thetabar)$ and $\epsilonhat \leq \epsilon$. Furthermore, if beliefs are in $N_{\epsilonhat}(\thetabar)$ for all stages, then the convergent belief must also be in $N_{\epsilonhat}(\thetabar) \subseteq N_{\thetaepfin}(\thetabar)$. Based on Assumptions \ref{as:uh} -- \ref{as:convergence}, Theorem \ref{theorem:convergence} provides that the sequence of strategies converges. Since $N_{\epsilonhat}(\thetabar) \subseteq N_{\epsilon'}(\thetabar)$, we know from \emph{(i)} that the convergent strategy is an equilibrium in the neighborhood $N_{\loadepfin}(\EQ(\thetabar))$. Thus, \emph{(iii)} holds.

Thanks to Lemma \ref{lemma:constrained_set} \emph{(iii)}, to prove local stability as in \eqref{eq:local}, it remains to be established that there exists an initial $(\epsilon^1, \delta^1)$-neighborhood of the fixed point such that when learning starts from that neighborhood, all states remain in the auxiliary $(\hat{\epsilon}, \delta)$- neighborhood with probability higher than $\gamma$. To characterize $\epsilon^1$, we note that $\thetat$ remains in the $\epsilonhat$-neighborhood of $\thetabar$ if $|\thetat(\s)-\thetabar(\s)|\leq \frac{\epsilonhat}{|\S|}$ for all $\s \in \S$. In Lemma \ref{lemma:stopping_time}, we characterize the condition of the initial belief $\thetazero$ under which $|\thetat(\s)-\thetabar(\s)|\leq \frac{\epsilonhat}{|\S|}$ with probability higher than $\gamma$ for all $\t$ and all $\s \in \S\setminus [\thetabar]$ (i.e. the set of parameters with zero probability in $\thetabar$). In Lemma \ref{lemma:other_parameters}, we analyze conditions of initial beliefs of parameters $s \in [\thetabar]$, which together with Lemma \ref{lemma:stopping_time} allow us to precisely characterize $\epsilon^1$. We also characterize $\delta^1$ that guarantees that the strategies do not leave $N_{\delta}(\EQ(\thetabar))$. Our characterization of $\epsilon^1$ and $\delta^1$ builds on the martingale property of belief ratios (Lemma \ref{lemma:thetatil}), and the fact that under Assumption \ref{as:stability}, states satisfy local consistency and local invariance in the auxiliary neighborhood (Lemma \ref{lemma:constrained_set}). 



Before proceeding, we need to define the following thresholds:

\begin{subequations}
\begin{align}
    0<\rhoone &< \min_{\s \in [\thetabar]} \left\{\frac{(1-\gamma) \thetabar(\s)\epsilonhat}{(1-\gamma +|\Sbar|)(|\S\setminus [\thetabar]|+1)|\S|+(1-\gamma) \epsilonhat}\right\}, \label{eq:rhoone} \\
    \rhotwo &\deleq\frac{\epsilonhat}{(|\Sbar|+1)|\S|}, \label{eq:rhotwo}\\
    0<\rhothree &< \min_{\s \in [\thetabar]}\left\{\frac{\epsilonhat - |\Sbar| |\S|\rhotwo \thetabar(\s)}{|\S|-|\Sbar||\S| \rhotwo},~   \frac{\hat{\epsilon}}{|S|} - \rhotwo |S\setminus [\thetabar]|\left(\thetabar(s)+ \frac{\hat{\epsilon}}{|S|}\right), ~ \thetabar(\s) \right\}. \label{eq:rho_three}
\end{align}
\end{subequations}


Lemma \ref{lemma:stopping_time} below shows that if the initial belief $\thetazero$ is in the neighborhood $N_{\rhoone}(\thetabar)$, then $\thetat(\s)< \rhotwo$ for all $\s \in \S\setminus [\thetabar]$ in all stages of the learning dynamics with probability higher than $\gamma$. Note that $\thetat(\s) < \rhotwo$ ensures $|\thetat(\s)-\thetabar(\s)|< \frac{\epsilonhat}{|\S|}$ since $\thetabar(\s)=0$ for all $\s \in \Sbar$ and $\rhotwo < \frac{\epsilonhat}{|\S|}$. Additionally, the threshold $\rhotwo$ is specifically constructed to bound the beliefs of the remaining parameters in $[\thetabar]$, which will be used later in Lemma \ref{lemma:other_parameters}.

\begin{lemma}\label{lemma:stopping_time}
For any $\gamma \in (0, 1)$, if the initial belief satisfies
\begin{subequations}
\begin{align}
    &\thetazero(\s)< \rhoone, \quad \forall \s \in \Sbar, \label{eq:epsilonlow}\\
    &\thetabar(\s) - \rhoone < \thetazero(\s) < \thetabar(\s)+\rhoone, \quad \forall \s \in [\thetabar],\label{eq:epsilonlow_positive}
\end{align}
\end{subequations}
then 
\begin{align}\label{eq:nonsupport}
    \pro\(\thetat(\s)< \rhotwo, ~\forall \s \in \Sbar, ~ \forall \t\)> \gamma.
\end{align}
\end{lemma}

In the proof of Lemma \ref{lemma:stopping_time}, we say that the belief $\thetat(\s)$ completes an upcrossing of the interval $[\rhoone, \rhotwo]$ if $\thetat(\s)$ increases from less than $\rhoone$ to higher than $\rhotwo$. Note that if the belief of a parameter $\s \in \Sbar$ is initially smaller than $\rhoone$ but later becomes higher than $\rhotwo$ in some stage $\t$, then the belief sequence $\(\theta^j(\s)\)_{j=1}^{\t}$ must have completed at least one upcrossing of $[\rhoone, \rhotwo]$ before stage $\t$. Therefore, $\thetat(\s)<\rhotwo$ for all $\t$ is equivalent to that the number of upcrossings completed by the belief is zero. 

Additionally, by bounding the initial belief of parameters $\s \in [\thetabar]$ as in \eqref{eq:epsilonlow_positive}, we construct another interval $\left[\rhoone/\(\thetabar(\sran)- \rhoone\), \rhotwo\right]$ such that the number of upcrossings with respect to this interval completed by the auxiliary sequence of belief ratios $\(\frac{\thetatilt(\s)}{\thetatilt(\sran)}\)_{\t=1}^{\infty}$ is no less than the number of upcrossings with respect to interval $[\rhoone, \rhotwo]$ completed by $\(\thetat(\s)\)_{\t=1}^{\infty}$. Recall that the sequence of belief ratios $\(\frac{\thetatilt(\s)}{\thetatilt(\sran)}\)_{\t=1}^{\infty}$ forms a martingale process (Lemma \ref{lemma:thetatil}). By applying Doob's upcrossing inequality, we obtain an upper bound on the expected number of upcrossings completed by the belief ratio corresponding to each parameter $\s \in \Sbar$, which is also an upper bound on the expected number of upcrossings made by the belief of $\s$. Using Markov's inequality and the upper bound of the expected number of upcrossings, we show that with probability higher than $\gamma$, no belief $\thetat(\s)$ of any parameter $\s \in \Sbar$ can ever complete a single upcrossing with respect to the interval $[\rhoone, \rhotwo]$ characterized by \eqref{eq:rhoone} -- \eqref{eq:rhotwo}. Hence, $\thetat(\s)$ remains lower than the threshold $\rhotwo$ for all $\s \in \Sbar$ and all $\t$ with probability higher than $\gamma$.

\medskip 

\noindent \emph{Proof of Lemma \ref{lemma:stopping_time}.} First, note that $0< \rhoone < \rhotwo< \frac{\epsilonhat}{|\S|}$. For any $\s \in \Sbar$ and any $\t>1$, we denote $U^{\t}(\s)$ the number of upcrossings of the interval $[\rhoone, \rhotwo]$ that the belief $\thetatil^j(\s)$ completes by stage $\t$. That is, $U^{\t}(\s)$ is the maximum number of intervals $\([\underline{\t}_{i}, \overline{\t}_{i}]\)_{\i=1}^{U^{\t}(\s)}$ with $1 \leq \underline{\t}_{1} < \overline{\t}_{1} < \underline{\t}_{2} <  \overline{\t}_{2}< \cdots< \underline{\t}_{U^{\t}(\s)}< \overline{\t}_{U^{\t}(\s)} \leq \t$, such that $\thetatil^{\underline{\t}_{i}}(\s)<\rhoone< \rhotwo <\thetatil^{\overline{\t}_{i}}(\s)$ for $i=1, \dots U^{\t}(\s)$. Since the beliefs $\(\thetatil^\j(\s)\)_{\j=1}^{\t}$ are updated based on randomly realized payoffs $\(\c^j\)_{j=1}^{\t}$ as in \eqref{eq:update_belief_always}, $U^{\t}(\s)$ is also a random variable. For any $\t>1$, $U^{\t}(\s)\geq1$ if and only if $\thetatil^1(\s)<\rhoone$ and there exists a stage $j \leq \t$ such that $\thetatil^j(\s)>\rhotwo$. Equivalently, $\lim_{\t \to \infty} U^\t(\s)\geq1$ if and only if $\thetatil^1(\s)<\rhoone$ and there exists a stage $\t>1$ such that $\thetatilt(\s)>\rhotwo$. Therefore, if $\thetatil^1(\s)< \rhoone$ for all $\s \in \Sbar$, then:
\begin{align}
    &\pro\(\thetatilt(\s)<\rhotwo, ~ \forall \s \in \Sbar, ~\forall \t\) = 1- \pro\(\exists \s \in \Sbar \text{ and } \t, ~s.t. ~\thetatilt(\s)>\rhotwo\) \notag\\
    \geq & 1- \sum_{\s \in \Sbar} \pro\(\exists \t, ~s.t. ~~\thetatilt(\s)>\rhotwo\) = 1- \sum_{\s \in 
    \Sbar}\lim_{\t \to \infty} \pro\(U^\t(\s)\geq 1\). \label{eq:up_one}
\end{align}

Next, we define $\alpha \deleq \thetabar(\sran)-\rhoone$. Since $0<\rhoone < \min_{\s \in [\thetabar]} \{\thetabar(\s)\}$ and $\sran$ is in the support set, we have $\alpha \in (0, \thetabar(\sran))$. If $\thetatil^1(\s)$ satisfies \eqref{eq:epsilonlow} -- \eqref{eq:epsilonlow_positive}, then $\frac{\thetatil^1(\s)}{\thetatil^1(\sran)}< \frac{\rhoone}{\alpha}$ for all $\s \in \Sbar$. Additionally, for any stage $\t$ and any $\s \in \Sbar$, if $\thetatilt(\s)>\rhotwo$, then $\frac{\thetatilt(\s)}{\thetatilt(\sran)}> \rhotwo$ because $\thetatilt(\sran)\leq 1$. Hence, whenever $\thetatilt(\s)$ completes an upcrossing of the interval $\left[\rhoone, \rhotwo\right]$, $\frac{\thetatilt(\s)}{\thetatilt(\sran)}$ must also have completed an upcrosssing of the interval $\left[\frac{\rhoone}{\alpha}, \rhotwo\right]$. To ensure that the interval $\left[\frac{\rhoone}{\alpha}, \rhotwo\right]$ is non-empty, we need to verify that $\frac{\rhoone}{\alpha} = \frac{\rhoone}{\thetabar(\sran)- \rhoone}< \rhotwo$, which requires that 
\begin{align}\label{eq:verify}
    \rhoone < \frac{\rhotwo \thetabar(\sran)}{1+\rhotwo}= \frac{\epsilonhat \thetabar(\sran)}{\epsilonhat + (|S\setminus[\thetabar]|+1)|S|}.
\end{align}
From \eqref{eq:rhoone}, since $\sran \in [\thetabar]$, we can check that \eqref{eq:verify} is indeed satisfied. Thus, the interval $\left[\frac{\rhoone}{\alpha}, \rhotwo\right]$ is non-empty. 

We denote $\Uhat^\t(\s)$ as the number of upcrossings of the sequence $\(\frac{\thetatil^j(\s)}{\thetatil^j(\sran)}\)_{j=1}^{\t}$ with respect to the interval $\left[\frac{\rhoone}{\alpha}, \rhotwo\right]$ until stage $\t$. Then, $U^\t(\s) \leq \Uhat^\t(\s)$ for all $\t$. Therefore, we can write:
\begin{align}\label{eq:up_two}
    \pro\(U^\t(\s)\geq 1\) \leq \pro\(\Uhat^\t(\s)\geq 1\) \leq \mathbb{E}\left[\Uhat^\t(\s)\right], 
\end{align}
where the last inequality follows from the Makov inequality. 

From the proof of Lemma \ref{lemma:theta}, we know that the sequence $\(\frac{\thetatilt(\s)}{\thetatilt(\sran)}\)_{\t=1}^{\infty}$ is a martingale. Therefore, we can apply the Doob's upcrossing inequality as follows: 
\begin{align}\label{eq:up_three}
    \mathbb{E}\left[\Uhat^\t(\s)\right] \leq \frac{\mathbb{E}\left[\max \{\frac{\rhoone}{\alpha}-\frac{\thetatilt(\s)}{\thetatilt(\sran)}, 0\}\right]}{\rhotwo- \frac{\rhoone}{\alpha}} \leq \frac{\frac{\rhoone}{\alpha}}{\rhotwo- \frac{\rhoone}{\alpha}}, \quad \forall \t.
\end{align}
From \eqref{eq:up_one} -- \eqref{eq:up_three}, and \eqref{eq:rhoone} -- \eqref{eq:rhotwo}, we have: 
\begin{align*}
     \pro\(\thetatilt(\s)<\rhotwo, ~ \forall \s \in \Sbar, ~ \forall \t\) \geq 1- \frac{\frac{\rhoone}{\alpha} |\Sbar|}{\rhotwo- \frac{\rhoone}{\alpha}}= 1- \frac{\frac{\rhoone}{\thetabar(\sran)-\rhoone} |\Sbar|}{\rhotwo- \frac{\rhoone}{\thetabar(\sran)-\rhoone}} > \gamma,
\end{align*} 
where the last inequality is derived using \eqref{eq:rhoone} -- \eqref{eq:rhotwo}. Finally, from \eqref{eq:relationship}, we know that $\pro\(\thetat(\s)<\rhotwo, ~ \forall \s \in \Sbar, ~ \forall \t\)  \geq \pro\(\thetatilt(\s)<\rhotwo, ~ \forall \s \in \Sbar, ~ \forall \t\) >\gamma$.
\QEDA

Furthermore, Lemma \ref{lemma:other_parameters} characterizes another set of conditions on the initial state to ensure that the beliefs of remaining parameters $\s \in [\thetabar]$ satisfy $|\thetat(\s)-\thetabar(\s)|<\frac{\epsilonhat}{|\S|}$, and the strategy $\qt \in N_{\delta}\(\EQ(\thetabar)\)$ for all $\t$ so long as $\thetat(\s)<\rhotwo$ for any parameter $\s \in \S \setminus [\thetabar]$. Recall that $\thetat(\s)<\rhotwo$ for all $\s \in \S\setminus [\thetabar]$ is satisfied with probability higher than $\gamma$ under the conditions in Lemma \ref{lemma:stopping_time}.

\begin{lemma}\label{lemma:other_parameters}
Under Assumption \ref{as:stability}, if $|\thetazero(\s)-\thetabar(\s)|<\rhothree$ for all $\s \in [\thetabar]$ and $\q^1 \in N_{\delta}\(\EQ(\thetabar)\)$, then 
\begin{align}\label{eq:ensure}
\pro\(\left.\begin{array}{l}
|\thetat(\s)-\thetabar(\s)|< \frac{\epsilonhat}{|\S|}, ~\forall \s \in [\thetabar], ~\forall \t\\
\text{and } \qt \in N_{\delta}\(\EQ(\thetabar)\), ~\forall \t
\end{array}\right\vert
\thetat(\s)< \rhotwo, ~\forall \s \in \Sbar, ~\forall \t\)=1.
\end{align}
\end{lemma}

\vspace{0.2cm}
\noindent \emph{Proof of Lemma \ref{lemma:other_parameters}.}
From Assumption \emph{(A3a)}, we know that $[\thetabar] \subseteq \Sequiv(\q^1)$ if $\q^1 \in N_{\delta}\(\EQ(\thetabar)\)$. Hence, $\phibar^\s(\c^1|\q^1) = \phibar^{\sran}(\c^1|\q^1)$ for any $\s \in [\thetabar]$ and any realized payoff $\c^1$. Therefore, 
\begin{align}\label{eq:two_one}
    \frac{\thetatil^{2}(\s)}{\thetatil^{2}(\sran)}=\frac{\thetatil^{1}(\s)}{\thetatil^{1}(\sran)}\frac{\phibar^{\s}(\c^1|\q^1)}{\phibar^{\sran}(\c^1|\q^1)}= \frac{\thetatil^{1}(\s)}{\thetatil^{1}(\sran)}, \quad w.p.~1, \quad \forall \s \in [\thetabar].
\end{align}
This implies that $\frac{\sum_{\s \in [\thetabar]} \thetatil^{2}(\s)}{\thetatil^{2}(\sran)}=\frac{\sum_{\s \in [\thetabar]} \thetatil^1(\s)}{\thetatil^1(\sran)}$, and for all $\s \in [\thetabar]$:
\begin{align*}
\frac{\thetatil^{2}(\s)}{\sum_{\s \in [\thetabar]} \thetatil^{2}(\s)}=\frac{\thetatil^{2}(\s)}{\thetatil^{2}(\sran)} \frac{\thetatil^{2}(\sran)}{\sum_{\s \in [\thetabar]} \thetatil^{2}(\s)}=\frac{\thetatil^1(\s)}{\thetatil^1(\sran)} \frac{\thetatil^1(\sran)}{\sum_{\s \in [\thetabar]} \thetatil^1(\s)}=\frac{\thetatil^1(\s)}{\sum_{\s \in[\thetabar]} \thetatil^1(\s)}.
\end{align*}
Thus, we have
\begin{align*}
    \frac{\thetatil^{2}(\s)}{\thetatil^1(\s)}=\frac{\sum_{\s \in [\thetabar]} \thetatil^{2}(\s)}{\sum_{\s \in[\thetabar]} \thetatil^1(\s)}, \quad w.p.~1,  \quad \forall \s \in [\thetabar].\end{align*}

Since $\sum_{\s \in [\thetabar]} \thetatil^1(\s) \leq 1$, and conditioned on that $\thetatil^{2}(\s)<\rhotwo$ for all $\s \in \Sbar$ as in \eqref{eq:ensure}, we have 
\begin{align*}
    \frac{\thetatil^{2}(\s)}{\thetatil^1(\s)} = \frac{\sum_{\s \in [\thetabar]} \thetatil^{2}(\s)}{\sum_{\s \in[\thetabar]} \thetatil^1(\s)} \geq \sum_{\s \in [\thetabar]} \thetatil^{2}(\s) > 1-|\Sbar|\rhotwo, \quad \forall s \in [\thetabar].
\end{align*} Additionally, since $\sum_{\s \in [\thetabar]} \thetatil^{2}(\s) \leq 1$ and again conditioned on that $\thetatil^1(\s)< \rhotwo$ for all $\s \in S\setminus [\thetabar]$ as in \eqref{eq:ensure}, we have 
\begin{align*}
    \frac{\thetatil^{2}(\s)}{\thetatil^1(\s)}= \frac{\sum_{\s \in [\thetabar]} \thetatil^{2}(\s)}{\sum_{\s \in[\thetabar]} \thetatil^1(\s)} \leq \frac{1}{\sum_{\s \in[\thetabar]} \thetatil^1(\s)}<\frac{1}{1-|\Sbar| \rhotwo}, \quad \forall s \in [\thetabar].
\end{align*} 
Since by \eqref{eq:rho_three}, $\rhothree <  \thetabar(\s)$ for all $\s \in [\thetabar]$, any $\thetatil^1(\s) \in \(\thetabar(\s)-\rhothree, \thetabar(\s)+ \rhothree\)$ is a positive number for all $\s \in [\thetabar]$. Therefore, we have the following bounds: 
\begin{align}\label{eq:refer_end}
\(\thetabar(\s)-\rhothree\) \(1-|\Sbar|\rhotwo\)<\thetatil^{2}(\s) <\frac{\thetabar(\s)+ \rhothree}{1-|\Sbar| \rhotwo}, \quad \forall s \in [\thetabar].
\end{align}
Since
\begin{align}\label{eq:up}
    \rhothree \stackrel{\eqref{eq:rho_three}}{<} \frac{\epsilonhat - |\Sbar| |\S|\rhotwo \thetabar(\s)}{|\S|-|\Sbar||\S| \rhotwo}, \quad \forall \s \in [\thetabar],
\end{align}
we can check that $\(\thetabar(\s)-\rhothree\) \(1-|\Sbar|\rhotwo\) > \thetabar(\s) -\frac{\epsilonhat}{|\S|}$ for all $\s \in [\thetabar]$. To ensure the right-hand-side of \eqref{eq:up} is positive, we need to have $\rhotwo < \frac{\epsilonhat}{|\Sbar| |\S| \thetabar(\s)}$ for all $\s \in [\thetabar]$, and $\rhotwo < 1/|\Sbar|$, which is indeed satisfied by \eqref{eq:rhotwo}.

Also, since 
\begin{align}\label{eq:check_two}\rhothree \stackrel{\eqref{eq:rho_three}}{<} \frac{\hat{\epsilon}}{|S|} - \rhotwo |S\setminus [\thetabar]|\left(\thetabar(s)+ \frac{\hat{\epsilon}}{|S|}\right), \quad \forall s \in [\thetabar].
\end{align}
we have $\frac{\thetabar(\s)+ \rhothree}{1-|\Sbar| \rhotwo} <  \thetabar(\s)+\frac{\epsilonhat}{|\S|}$ for all $\s \in [\thetabar]$. Below, we check that the right-hand-side of \eqref{eq:check_two} is positive, and thus the constraint of $\rhothree$ in \eqref{eq:check_two} is feasible: 
\begin{align*}
    \frac{\hat{\epsilon}}{|S|} - \rhotwo |S\setminus [\thetabar]|\left(\thetabar(s)+ \frac{\hat{\epsilon}}{|S|}\right) &\geq \frac{\hat{\epsilon}}{|S|} - \rhotwo |S\setminus [\thetabar]|\left(1+ \frac{\hat{\epsilon}}{|S|}\right) \stackrel{\eqref{eq:rhotwo}}{=} \frac{\hat{\epsilon}}{|S|} \frac{1}{|S\setminus [\thetabar]|+1}- \frac{\hat{\epsilon}}{|S|} \frac{\epsilonhat |S\setminus [\thetabar]|}{(|\Sbar|+1)|\S|}\\
    &= \frac{\hat{\epsilon}}{|S|} \frac{|S|- \hat{\epsilon}|S\setminus [\thetabar]|}{(|\Sbar|+1)|\S|}>0.
\end{align*}From \eqref{eq:refer_end}, we can conclude that $|\thetatil^{2}(\s) - \thetabar(s)|< \frac{\epsilonhat}{|\S|}$ for all $\s \in [\thetabar]$. Additionally, conditional on $\thetatil^2(\s) <\rhotwo < \frac{\epsilonhat}{|\S|}$ for all $\s \in \S \setminus [\thetabar]$ as in \eqref{eq:ensure}, we have $\thetatil^2 \in N_{\epsilonhat}\(\thetabar\)$. From \eqref{eq:relationship}, we have $\theta^2 \in N_{\epsilonhat}\(\thetabar\)$. From \emph{(ii)} in Lemma \ref{lemma:constrained_set}, since $\q^1 \in N_{\delta}\(\EQ(\thetabar)\)$, we know that the updated strategy $q^2= F(\theta^2,\q^1) \in N_{\delta}\(\EQ(\thetabar)\)$. 

\smallskip

We now use mathematical induction to prove that the belief of any $\s \in [\thetabar]$ satisfies $|\thetatil^{k}(\s) - \thetabar(s)| <\frac{\epsilonhat}{|\S|}$ for stages $\t\geq 2$. If in stages $j=1, \dots, \t$, $|\thetatil^j(\s)-\thetabar(\s)| < \frac{\epsilonhat}{|\S|}$ for all $\s \in [\thetabar]$ and $\thetatil^j(\s)<\rhotwo<\frac{\epsilonhat}{|\S|}$ for all $\s \in \Sbar$, then $\thetatil^j \in N_{\epsilonhat}\(\thetabar\)$ for all $j=1, \dots, \t$. Thus, from \eqref{eq:relationship} and \emph{(ii)} in Lemma \ref{lemma:constrained_set}, we have $\theta^j \in N_{\epsilonhat}\(\thetabar\)$ and $F\(\theta^j, \q^{j-1}\) \subseteq N_{\delta}\(\EQ(\thetabar)\)$. Therefore, $\qj \in N_{\delta}\(\EQ(\thetabar)\)$ for all $j=2, \dots, \t$. 

\smallskip
From Assumption \emph{(A3a)}, we know that  $[\thetabar] \subseteq \Sequiv(\qj)$ for all $j=1, \dots, \t$. Therefore, for any $\s \in [\thetabar]$ and any $j=1, \dots, \t$, $\phibar^\s(\cj|\qj)=\phibar^{\sran}(\cj|\qj)$ with probability 1. Then, by iteratively applying \eqref{eq:two_one} for $j=1, 2, \dots k$, we have $\frac{\thetatil^{\t+1}(\s)}{\thetatil^1(\s)}=\frac{\sum_{\s \in [\thetabar]} \thetatil^{\t+1}(\s)}{\sum_{\s \in[\thetabar]} \thetatil^1(\s)}$ for all $\s \in [\thetabar]$ with probability 1. Analogous to $\t=2$, we can prove that $|\thetatil^{\t+1}(\s)-\thetabar(\s)|< \frac{\epsilonhat}{|\S|}$ for all $\s \in [\thetabar]$. From \eqref{eq:relationship}, we also have $|\theta^{\t+1}(\s)-\thetabar(\s)|<\frac{\epsilonhat}{|\S|}$ for all $\s \in [\thetabar]$. From the principle of mathematical induction, we conclude that in all stages $\t$,  $|\thetat(\s)-\thetabar(\s)|<  \frac{\epsilonhat}{|\S|}$ for all $\s \in [\thetabar]$, and $\qt \in N_{\delta}\(\EQ(\thetabar)\)$ for all $\t$. Therefore, we have proved \eqref{eq:ensure}.

\QEDA

\vspace{0.2cm}

Finally, for any $\gamma \in (0, 1)$, and any $\thetaepfin, \loadepfin >0$, we set $\thetaep = \min\{\rhoone, \rhothree\}$ given by \eqref{eq:rhoone} -- \eqref{eq:rho_three}, and $\loadep =\delta$. We are now ready to prove Theorem \ref{theorem:stability}. 

\vspace{0.2cm}
\noindent\emph{Proof of Theorem \ref{theorem:stability}.}
If $\thetazero \in N_{\epsilon^1}(\thetabar)$, then $|\thetazero(\s)-\thetabar(\s)|< \thetaep$ for all $\s \in \S$. Recall from \emph{(iii)} in Lemma \ref{lemma:constrained_set}, $\lim_{\t \to \infty} \pro\(\thetat\in N_{\thetaepfin}(\theta), ~ \qt \in \neighinfload(\EQ(\thetabar)) \) \geq  \pro\(\thetat \in N_{\epsilonhat}(\thetabar), ~ \qt \in N_{\delta}\(\EQ(\thetabar)\), ~ \forall \t\)$. Since $\rhotwo < \epsilonhat/|\S|$, we further have: 
\begin{align*}
    &\pro\(\thetat \in N_{\epsilonhat}(\thetabar), ~ \qt \in N_{\delta}\(\EQ(\thetabar)\), ~ \forall \t\) \geq \pro\(\begin{array}{l}
|\thetat(\s)-\thetabar(\s)|< \frac{\epsilonhat}{|\S|}, ~ \forall \s \in [\thetabar],~ \forall \t \text{ and}\\
\thetat < \rhotwo,~ \forall \s \in \Sbar,~ \qt \in N_{\delta}\(\EQ(\thetabar)\), ~\forall \t
\end{array}\)\\
&= \pro\(\thetat(\s) < \rhotwo, \forall \s \in \Sbar, \forall \t\) \cdot \pro\(\left.\begin{array}{l}
|\thetat(\s)-\thetabar(\s)|< \frac{\epsilonhat}{|\S|}, \forall \s \in [\thetabar], \forall \t\\
\text{and }\qt \in N_{\delta}\(\EQ(\thetabar)\), ~\forall \t
\end{array}\right\vert
\begin{array}{l}
\thetat(\s)< \rhotwo.\\
\forall \s \in \Sbar, \forall \t\end{array}\)
\end{align*}
For any $\thetazero \in N_{\epsilon^1}(\thetabar)$ and any $\q^1 \in N_{\loadep}\(\EQ(\thetabar)\)$, we know from Lemmas \ref{lemma:stopping_time} -- \ref{lemma:other_parameters} that: 
\begin{align*}
&\pro\(\thetat(\s)< \rhotwo, ~\forall \s \in \Sbar, ~\forall \t\)> \gamma, \text{ and } \\
&\pro\(\left.\begin{array}{l}
|\thetat(\s)-\thetabar(\s)|< \frac{\epsilonhat}{|\S|}, ~ \forall \s \in [\thetabar], ~\forall \t\\
\text{and }\qt \in N_{\delta}\(\EQ(\thetabar)\), ~\forall \t
\end{array}\right\vert
\thetat(\s)< \rhotwo, ~\forall \s \in \Sbar, ~\forall \t\) =1
\end{align*}
Therefore, for any $\thetazero \in N_{\epsilon^1}(\thetabar)$ and any $\q^1 \in N_{\loadep}\(\EQ(\thetabar)\)$, the states of learning dynamics satisfy $\lim_{\t \to \infty} \pro\(\thetat\in N_{\thetaepfin}(\theta), ~ \qt \in  N_{\loadepfin}\(\EQ(\thetabar)\) \)> \gamma$. Thus, $\(\thetabar, \qbar\)$ is locally stable.
\QEDA




    
\section{Extensions}\label{sec:variant}
In this section, we briefly discuss two extensions of the learning model introduced in Section \ref{sec:basic_model}: (1) Learning with two timescales; (2) Learning in games with finite strategies.

\vspace{0.2cm}

\noindent\textbf{(1) Learning with two timescales.} Consider the case where strategy update is at a faster timescale compared with the belief updates, i.e. $\lim_{t \to \infty}\t_{t+1}-\kt = \infty$ with probability 1. Then, our convergence and stability results still hold as follows: 
\begin{proposition}\label{prop:timescale}
Under Assumption \ref{as:convergence}, the sequence of states $\(\thetat, \qt\)_{\t=1}^{\infty}$ induced by \eqref{eq:update_belief} -- \eqref{eq:generic} converges to a fixed point $\(\thetabar, \qbar\)$ with probability 1, and $\(\thetabar, \qbar\)$ satisfies \eqref{eq:exclude_distinguished} and \eqref{subeq:eq_fixed}. 

Additionally, under Assumption \ref{as:convergence}, a fixed point $(\thetabar, \qbar)$ is locally stable if Assumption \ref{as:stability} is satisfied. 
\end{proposition}

We note that the convergence and stability results in Proposition \ref{prop:timescale} does not rely on Assumption \ref{as:uh}. Since $\lim_{t \to \infty}\t_{t+1}-\kt = \infty$ with probability 1, as $t \rightarrow \infty$, the strategies between any two belief updates $\kt$ and $\t_{t+1}$ are updated with the static belief $\theta^{\kt}$. Therefore, Assumption \ref{as:convergence} directly ensures that strategies converge to an equilibrium strategy profile in $\EQ\(\theta^{\kt}\)$ without relying on the construction of auxiliary strategies or Assumption \ref{as:uh}. Then, the updated belief $\theta^{\t_{t+1}}$ will form an accurate payoff estimate given the equilibrium strategy following Lemma \ref{lemma:consistency}. By repeating this process, we can show that the beliefs and strategies converge to a fixed point. Similarly, the local stability result does not rely on Assumption \ref{as:uh}, and it holds following Theorem \ref{theorem:stability}. The global stability result is also analogous to that in Proposition \ref{prop:global}. 

\vspace{0.2cm}
\noindent\textbf{(2) Learning in Games with Finite Strategy Set.}
Our results in Section \ref{sec:main} can be extended to learning in games where strategy sets are finite and players can choose mixed strategies. In this game, each player $\i$'s action set (pure strategies) is a finite set $\Ai$, and the action profile (pure strategy profile) is denoted as $\a=\(\ai\)_{\i \in \I} \in A= \prod_{\i \in \I} \Ai$. Given any parameter $\s$ and any action profile $\a$, the distribution of players' payoff $\c$ is $\phi^\s\(\c|\a\)$.  

We denote player $\i$'s mixed strategy as $\qi = \(\qi(\ai)\)_{\ai \in \Ai} \in \Qi = \Delta\(\Ai\)$, where $\qi(\ai)$ is the probability of choosing the action $\ai$. The mixed strategy set $\Qi$ is bounded and convex. Players' action profile in each stage $\t$, denoted as $\actiont =\(\ai^{\t}\)_{\i \in \I}$, is realized from the mixed strategy profile $\qt$. Analogous to \eqref{eq:update_belief} -- \eqref{eq:generic}, beliefs and strategies in each stage $\t$ are updated based on the past actions $\(\a^{j}\)_{j=1}^{\t}$ and the realized payoff vectors $\(\c^j\)_{j=1}^{\t}$.  

The convergence result in Theorem \ref{theorem:convergence} can be readily extended to games with finite strategy sets: Under Assumptions \ref{as:uh} and \ref{as:convergence}, the sequence of beliefs and mixed strategies $\(\thetat, \qt\)_{\t =1}^{\infty}$ converges to a fixed point $(\thetabar, \qbar)$, where $\qbar \in \EQ(\thetabar)$ is a mixed Nash equilibrium associated with the belief $\thetabar$, and $\thetabar$ accurately estimates the payoff distribution for all action profiles that are taken with positive probability in $\qbar$. Additionally, our results on global and local stability properties (Proposition \ref{prop:global} and Theorem \ref{theorem:stability}) also hold analogously.

Moreover, in games with finite strategy sets, any local neighborhood of a fixed point strategy profile includes mixed strategies with full support on all action profiles, and any parameter $\s \neq \sran$ can be distinguished from $\sran$ with the realized payoffs. In this case, only complete information fixed points satisfy the local consistency condition \emph{(A3a)}, and thus is locally stable (Proposition \ref{prop:complete_local}). Any other fixed point is non-robust to local perturbation of strategies. Hence, our stability analysis implies that complete learning is guaranteed by local exploration in games with finite strategies.\footnote{Our observation that local exploration is sufficient to identify the complete information equilibrium in games with finite strategies is consistent with literature on equilibrium learning in normal form games with unknown payoff matrices (\citet{hart2003regret, foster2006regret, marden2007regret, daskalakis2011near, syrgkanis2015fast}). Indeed, algorithms proposed in these literature rely on local exploration -- repeatedly sampling actions that are not equilibrium  with small probability to learn the payoff matrices. }

  \section{Concluding Remarks}
In this article, we studied stochastic learning dynamics induced by a set of strategic players who repeatedly play a game with an unknown parameter. We analyzed the convergence of beliefs and strategies induced by the stochastic dynamics, and derived conditions for local and global stability of fixed points. We also provided conditions that guarantee that learning converges to a complete information equilibrium. 

A future research question of interest is to analyze the learning dynamics when players seek to efficiently learn the true parameter by choosing off-equilibrium strategies. When there are one or more parameters that are payoff equivalent to the true parameter at fixed point, complete learning requires players to take strategies that may reduce their individual payoffs in some stages. In our setup, if a player were to choose a non-equilibrium strategy, the information resulting from that player's realized payoff would be incorporated into the belief update, and the new belief is known to all players. Under what scenarios the utility-maximizing players will choose their strategies to engage such explorative behavior is an interesting question, and worthy of further investigation. 

Another promising extension is to study multi-agent reinforcement learning problem from a Bayesian viewpoint. In such settings, the unknown parameter changes over time according to a Markovian transition process, and players may have imperfect or no knowledge of the underlying transition kernel. The approach presented in this article on studying the dynamic interplay between belief updates and strategy updates is useful to analyze how players learn the belief estimates of payoffs that depend on the latent Markov state, and adaptively adjust their strategies that converges to a stationary equilibrium. 

\section*{Acknowledgement}
The authors are grateful to the area editor, the associate editor and the three reviewers for useful suggestions and constructive feedback. We thank participants of the Cornell ORIE Colloquium (2021), Games, Decisions and Networks Seminar (2021), seminar at Simons Institute for the Theory of Computing (2022), and the Spring 2022 Colloquium at the C3.ai Digital Transformation Institute for useful discussions. This research was supported in part by Simons Fellowship, Michael Hammer Fellowship, and AFOSR project Building Attack Resilience into Complex Networks.  
\bibliographystyle{plainnat}
\bibliography{library.bib}

\begin{thebibliography}{48}
\providecommand{\natexlab}[1]{#1}
\providecommand{\url}[1]{\texttt{#1}}
\expandafter\ifx\csname urlstyle\endcsname\relax
  \providecommand{\doi}[1]{doi: #1}\else
  \providecommand{\doi}{doi: \begingroup \urlstyle{rm}\Url}\fi

\bibitem[Acemoglu et~al.(2014)Acemoglu, Bimpikis, and
  Ozdaglar]{acemoglu2014dynamics}
Daron Acemoglu, Kostas Bimpikis, and Asuman Ozdaglar.
\newblock Dynamics of information exchange in endogenous social networks.
\newblock \emph{Theoretical Economics}, 9\penalty0 (1):\penalty0 41--97, 2014.

\bibitem[Acemoglu et~al.(2017)Acemoglu, Makhdoumi, Malekian, and
  Ozdaglar]{acemoglu2017fast}
Daron Acemoglu, Ali Makhdoumi, Azarakhsh Malekian, and Asuman Ozdaglar.
\newblock Fast and slow learning from reviews.
\newblock Technical report, National Bureau of Economic Research, 2017.

\bibitem[Al{\'o}s-Ferrer and Netzer(2010)]{alos2010logit}
Carlos Al{\'o}s-Ferrer and Nick Netzer.
\newblock The logit-response dynamics.
\newblock \emph{Games and Economic Behavior}, 68\penalty0 (2):\penalty0
  413--427, 2010.

\bibitem[Auer et~al.(2002)Auer, Cesa-Bianchi, and Fischer]{auer2002finite}
Peter Auer, Nicolo Cesa-Bianchi, and Paul Fischer.
\newblock Finite-time analysis of the multiarmed bandit problem.
\newblock \emph{Machine learning}, 47\penalty0 (2):\penalty0 235--256, 2002.

\bibitem[Banerjee(1992)]{banerjee1992simple}
Abhijit~V Banerjee.
\newblock A simple model of herd behavior.
\newblock \emph{The Quarterly Journal of Economics}, 107\penalty0 (3):\penalty0
  797--817, 1992.

\bibitem[Beggs(2005)]{beggs2005convergence}
Alan~W Beggs.
\newblock On the convergence of reinforcement learning.
\newblock \emph{Journal of Economic Theory}, 122\penalty0 (1):\penalty0 1--36,
  2005.

\bibitem[Bena{\i}m and Hirsch(1999)]{benaim1999mixed}
Michel Bena{\i}m and Morris~W Hirsch.
\newblock Mixed equilibria and dynamical systems arising from fictitious play
  in perturbed games.
\newblock \emph{Games and Economic Behavior}, 29\penalty0 (1-2):\penalty0
  36--72, 1999.

\bibitem[Blume et~al.(1993)]{blume1993statistical}
Lawrence~E Blume et~al.
\newblock The statistical mechanics of strategic interaction.
\newblock \emph{Games and Economic Behavior}, 5\penalty0 (3):\penalty0
  387--424, 1993.

\bibitem[Bravo et~al.(2018)Bravo, Leslie, and Mertikopoulos]{bravo2018bandit}
Mario Bravo, David Leslie, and Panayotis Mertikopoulos.
\newblock Bandit learning in concave n-person games.
\newblock \emph{Advances in Neural Information Processing Systems}, 31, 2018.

\bibitem[Cesa-Bianchi and Lugosi(2006)]{cesa2006prediction}
Nicolo Cesa-Bianchi and G{\'a}bor Lugosi.
\newblock \emph{Prediction, learning, and games}.
\newblock Cambridge University Press, 2006.

\bibitem[Cominetti et~al.(2010)Cominetti, Melo, and Sorin]{cominetti2010payoff}
Roberto Cominetti, Emerson Melo, and Sylvain Sorin.
\newblock A payoff-based learning procedure and its application to traffic
  games.
\newblock \emph{Games and Economic Behavior}, 70\penalty0 (1):\penalty0 71--83,
  2010.

\bibitem[Daskalakis et~al.(2011)Daskalakis, Deckelbaum, and
  Kim]{daskalakis2011near}
Constantinos Daskalakis, Alan Deckelbaum, and Anthony Kim.
\newblock Near-optimal no-regret algorithms for zero-sum games.
\newblock In \emph{Proceedings of the Twenty-second Annual ACM-SIAM Symposium
  on Discrete Algorithms}, pages 235--254. SIAM, 2011.

\bibitem[Dumett and Cominetti(2018)]{dumett2018stability}
Miguel~A Dumett and Roberto Cominetti.
\newblock On the stability of an adaptive learning dynamics in traffic games.
\newblock \emph{arXiv preprint arXiv:1807.01256}, 2018.

\bibitem[Foster and Young(2006)]{foster2006regret}
Dean Foster and Hobart~Peyton Young.
\newblock Regret testing: {L}earning to play {N}ash equilibrium without knowing
  you have an opponent.
\newblock \emph{Theoretical Economics}, 1\penalty0 (3):\penalty0 341--367,
  2006.

\bibitem[Fudenberg and Kreps(1993)]{fudenberg1993learning}
Drew Fudenberg and David~M Kreps.
\newblock Learning mixed equilibria.
\newblock \emph{Games and Economic Behavior}, 5\penalty0 (3):\penalty0
  320--367, 1993.

\bibitem[Fudenberg and Kreps(1995)]{fudenberg1995learning}
Drew Fudenberg and David~M Kreps.
\newblock Learning in extensive-form games {I}. self-confirming equilibria.
\newblock \emph{Games and Economic Behavior}, 8\penalty0 (1):\penalty0 20--55,
  1995.

\bibitem[Fudenberg and Levine(1993)]{fudenberg1993self}
Drew Fudenberg and David~K Levine.
\newblock Self-confirming equilibrium.
\newblock \emph{Econometrica: Journal of the Econometric Society}, pages
  523--545, 1993.

\bibitem[Fudenberg and Tirole(1991)]{fudenberg1991game}
Drew Fudenberg and Jean Tirole.
\newblock \emph{Game theory}.
\newblock MIT press, 1991.

\bibitem[Gale and Kariv(2003)]{gale2003bayesian}
Douglas Gale and Shachar Kariv.
\newblock {Bayesian} learning in social networks.
\newblock \emph{Games and Economic Behavior}, 45\penalty0 (2):\penalty0
  329--346, 2003.

\bibitem[Golowich et~al.(2020)Golowich, Pattathil, and
  Daskalakis]{golowich2020tight}
Noah Golowich, Sarath Pattathil, and Constantinos Daskalakis.
\newblock Tight last-iterate convergence rates for no-regret learning in
  multi-player games.
\newblock \emph{Advances in neural information processing systems},
  33:\penalty0 20766--20778, 2020.

\bibitem[Golub and Jackson(2010)]{golub2010naive}
Benjamin Golub and Matthew~O Jackson.
\newblock Naive learning in social networks and the wisdom of crowds.
\newblock \emph{American Economic Journal: Microeconomics}, 2\penalty0
  (1):\penalty0 112--49, 2010.

\bibitem[Hart and Mas-Colell(2000)]{hart2000simple}
Sergiu Hart and Andreu Mas-Colell.
\newblock A simple adaptive procedure leading to correlated equilibrium.
\newblock \emph{Econometrica}, 68\penalty0 (5):\penalty0 1127--1150, 2000.

\bibitem[Hart and Mas-Colell(2003)]{hart2003regret}
Sergiu Hart and Andreu Mas-Colell.
\newblock Regret-based continuous-time dynamics.
\newblock \emph{Games and Economic Behavior}, 45\penalty0 (2):\penalty0
  375--394, 2003.

\bibitem[Hofbauer and Sandholm(2002)]{hofbauer2002global}
Josef Hofbauer and William~H Sandholm.
\newblock On the global convergence of stochastic fictitious play.
\newblock \emph{Econometrica}, 70\penalty0 (6):\penalty0 2265--2294, 2002.

\bibitem[Hofbauer and Sandholm(2009)]{hofbauer2009stable}
Josef Hofbauer and William~H Sandholm.
\newblock Stable games and their dynamics.
\newblock \emph{Journal of Economic Theory}, 144\penalty0 (4):\penalty0
  1665--1693, 2009.

\bibitem[Hofbauer and Sorin(2006)]{hofbauer2006best}
Josef Hofbauer and Sylvain Sorin.
\newblock Best response dynamics for continuous zero-sum games.
\newblock \emph{Discrete and Continuous Dynamical Systems Series B}, 6\penalty0
  (1):\penalty0 215, 2006.

\bibitem[Hopkins(2002)]{hopkins2002two}
Ed~Hopkins.
\newblock Two competing models of how people learn in games.
\newblock \emph{Econometrica}, 70\penalty0 (6):\penalty0 2141--2166, 2002.

\bibitem[Lattimore and Szepesv{\'a}ri(2020)]{lattimore2020bandit}
Tor Lattimore and Csaba Szepesv{\'a}ri.
\newblock \emph{Bandit algorithms}.
\newblock Cambridge University Press, 2020.

\bibitem[Marden and Shamma(2012)]{marden2012revisiting}
Jason~R Marden and Jeff~S Shamma.
\newblock Revisiting log-linear learning: Asynchrony, completeness and
  payoff-based implementation.
\newblock \emph{Games and Economic Behavior}, 75\penalty0 (2):\penalty0
  788--808, 2012.

\bibitem[Marden et~al.(2007)Marden, Arslan, and Shamma]{marden2007regret}
Jason~R Marden, G{\"u}rdal Arslan, and Jeff~S Shamma.
\newblock Regret based dynamics: convergence in weakly acyclic games.
\newblock In \emph{Proceedings of the 6th International Joint Conference on
  Autonomous Agents and Multiagent Systems}, page~42. ACM, 2007.

\bibitem[Marden et~al.(2009)Marden, Young, Arslan, and
  Shamma]{marden2009payoff}
Jason~R Marden, H~Peyton Young, G{\"u}rdal Arslan, and Jeff~S Shamma.
\newblock Payoff-based dynamics for multiplayer weakly acyclic games.
\newblock \emph{SIAM Journal on Control and Optimization}, 48\penalty0
  (1):\penalty0 373--396, 2009.

\bibitem[Matsui(1992)]{matsui1992best}
Akihiko Matsui.
\newblock Best response dynamics and socially stable strategies.
\newblock \emph{Journal of Economic Theory}, 57\penalty0 (2):\penalty0
  343--362, 1992.

\bibitem[Meigs et~al.(2017)Meigs, Parise, and Ozdaglar]{meigs2017learning}
Emily Meigs, Francesca Parise, and Asuman Ozdaglar.
\newblock Learning dynamics in stochastic routing games.
\newblock In \emph{2017 55th Annual Allerton Conference on Communication,
  Control, and Computing (Allerton)}, pages 259--266. IEEE, 2017.

\bibitem[Mertikopoulos and Zhou(2019)]{mertikopoulos2019learning}
Panayotis Mertikopoulos and Zhengyuan Zhou.
\newblock Learning in games with continuous action sets and unknown payoff
  functions.
\newblock \emph{Mathematical Programming}, 173\penalty0 (1):\penalty0 465--507,
  2019.

\bibitem[Milgrom and Roberts(1990)]{milgrom1990rationalizability}
Paul Milgrom and John Roberts.
\newblock Rationalizability, learning, and equilibrium in games with strategic
  complementarities.
\newblock \emph{Econometrica: Journal of the Econometric Society}, pages
  1255--1277, 1990.

\bibitem[Moe and Fader(2004)]{moe2004dynamic}
Wendy~W Moe and Peter~S Fader.
\newblock Dynamic conversion behavior at e-commerce sites.
\newblock \emph{Management Science}, 50\penalty0 (3):\penalty0 326--335, 2004.

\bibitem[Monderer and Shapley(1996{\natexlab{a}})]{monderer1996fictitious}
Dov Monderer and Lloyd~S Shapley.
\newblock Fictitious play property for games with identical interests.
\newblock \emph{Journal of Economic Theory}, 68\penalty0 (1):\penalty0
  258--265, 1996{\natexlab{a}}.

\bibitem[Monderer and Shapley(1996{\natexlab{b}})]{monderer1996potential}
Dov Monderer and Lloyd~S Shapley.
\newblock Potential games.
\newblock \emph{Games and Economic Behavior}, 14\penalty0 (1):\penalty0
  124--143, 1996{\natexlab{b}}.

\bibitem[Mossel et~al.(2015)Mossel, Sly, and Tamuz]{mossel2015strategic}
Elchanan Mossel, Allan Sly, and Omer Tamuz.
\newblock Strategic learning and the topology of social networks.
\newblock \emph{Econometrica}, 83\penalty0 (5):\penalty0 1755--1794, 2015.

\bibitem[Rosen(1965)]{rosen1965existence}
J~Ben Rosen.
\newblock Existence and uniqueness of equilibrium points for concave n-person
  games.
\newblock \emph{Econometrica: Journal of the Econometric Society}, pages
  520--534, 1965.

\bibitem[Samuelson and Zhang(1992)]{samuelson1992evolutionary}
Larry Samuelson and Jianbo Zhang.
\newblock Evolutionary stability in asymmetric games.
\newblock \emph{Journal of Economic Theory}, 57\penalty0 (2):\penalty0
  363--391, 1992.

\bibitem[Sandholm(2010{\natexlab{a}})]{sandholm2010local}
William~H Sandholm.
\newblock Local stability under evolutionary game dynamics.
\newblock \emph{Theoretical Economics}, 5\penalty0 (1):\penalty0 27--50,
  2010{\natexlab{a}}.

\bibitem[Sandholm(2010{\natexlab{b}})]{sandholm2010population}
William~H Sandholm.
\newblock \emph{Population games and evolutionary dynamics}.
\newblock MIT press, 2010{\natexlab{b}}.

\bibitem[Smith and Price(1973)]{smith1973logic}
J~Maynard Smith and George~R Price.
\newblock The logic of animal conflict.
\newblock \emph{Nature}, 246\penalty0 (5427):\penalty0 15--18, 1973.

\bibitem[Syrgkanis et~al.(2015)Syrgkanis, Agarwal, Luo, and
  Schapire]{syrgkanis2015fast}
Vasilis Syrgkanis, Alekh Agarwal, Haipeng Luo, and Robert~E Schapire.
\newblock Fast convergence of regularized learning in games.
\newblock \emph{Advances in Neural Information Processing Systems}, 28, 2015.

\bibitem[Taylor and Jonker(1978)]{taylor1978evolutionary}
Peter~D Taylor and Leo~B Jonker.
\newblock Evolutionary stable strategies and game dynamics.
\newblock \emph{Mathematical Siosciences}, 40\penalty0 (1-2):\penalty0
  145--156, 1978.

\bibitem[Wu et~al.(2021)Wu, Amin, and Ozdaglar]{wu2021value}
Manxi Wu, Saurabh Amin, and Asuman~E Ozdaglar.
\newblock Value of information in {Bayesian} routing games.
\newblock \emph{Operations Research}, 69\penalty0 (1):\penalty0 148--163, 2021.

\bibitem[Zhu et~al.(2010)Zhu, Levinson, Liu, and Harder]{zhu2010traffic}
Shanjiang Zhu, David Levinson, Henry~X Liu, and Kathleen Harder.
\newblock {The traffic and behavioral effects of the I-35W Mississippi River
  Bridge collapse}.
\newblock \emph{Transportation Research Part A: Policy and Practice},
  44\penalty0 (10):\penalty0 771--784, 2010.

\end{thebibliography}

\appendix

\section{Supplementary Proofs}\label{apx:proof}


\noindent\emph{Proof of Lemma \ref{lemma:thetatil}.} 
Starting from any initial belief $\thetatil^1$, consider a sequence of strategies $Q^{\t-1}\deleq \(\qj\)_{j=1}^{\t-1}$ and a sequence of realized payoffs $Y^{\t-1}\deleq \(\cj\)_{\j =1}^{\t-1}$ before stage $\t$. The belief $\thetatil^{\t}$ is the repeatedly updated belief from $\thetatil^1$ based on $Q^{\t-1}$ and $Y^{\t-1}$ using \eqref{eq:update_belief_always}.  Then, from \eqref{eq:ratio}, we obtain the follows: 
\begin{align}\label{conditional_iterate}
\mathbb{E}\left[\left.\frac{\thetatil^{\t+1}(\s)}{\thetatil^{\t+1}(\sran)}\right\vert \thetatil^1, Q^{\t-1}, Y^{\t-1}\right]&= \frac{\thetatil^{\t}(\s)}{\thetatil^{\t}(\sran)} \cdot \mathbb{E}\left[ \frac{\phibar^{\s}(\ct|\qt)}{ \phibar^{\sran}(\ct|\qt)} \right],
\end{align}
where $\tilde{\theta}^{\t}$ and $\qt$ are derived from the belief and strategy updates given by $Q^{\t-1}$ and $Y^{\t-1}$. Note that
\begin{align*}
&\mathbb{E}\left[ \frac{\phibar^{\s}(\ct|\qt)}{ \phibar^{\sran}(\ct|\qt)}\right]= \int_{y^{\t}}  \(\frac{\phibar^{\s}(\ct|\qt)}{ \phibar^{\sran}(\ct|\qt)}\) \cdot  \phibar^{\sran}(\ct|\qt) d y^{\t}
=\int_{y^{\t}}  \phibar^{\s}(\ct|\qt) d y^{\t} =1.
\end{align*}
Hence, for any $\t=1, 2, \dots$,  
\begin{align*}
\mathbb{E}\left[\left.\frac{\thetatil^{\t+1}(\s)}{\thetatil^{\t+1}(\sran)}\right\vert \thetatil^1, Q^{\t-1}, Y^{\t-1}\right]&= \frac{\thetatil^{\t}(\s)}{\thetatil^{\t}(\sran)}, \quad \forall \s \in \S.
\end{align*}
Since $\frac{\thetatil^{\t}(\s)}{\thetatil^{\t}(\sran)}\geq 0$, the sequence $\(\frac{\thetatil^{\t}(\s)}{\thetatil^{\t}(\sran)}\)_{t=1}^{\infty}$ is a non-negative martingale for any $\s \in \S$. 

\medskip 
\noindent\emph{Proof of Lemma \ref{lemma:theta}.} From the martingale convergence theorem and Lemma \ref{lemma:thetatil}, we know that $\frac{\thetatil^{\t}(\s)}{\thetatil^{\t}(\sran)}$ converges with probability 1. Therefore, $\frac{\sum_{\s \in \S}\thetatil^{\t}(\s)}{\thetatil^{\t}(\sran)}$ converges with probability 1. Since $\sum_{\s \in \S}\thetatil^{\t}(\s)=1$ and $\thetatil^{\t}(\sran)>0$ for all $\t$, $\thetatil^{\t}(\sran)$ also converges with probability 1. We can thus conclude that $\thetatil^{\t}$ converges with probability 1 for all $s \in S$, and the convergent vector $\thetabar$ is a belief vector. Moreover, from \eqref{eq:relationship}, we know that $\thetat$ also converges to $\thetabar$ with probability 1. \QEDA

\vspace{0.2cm}
\begin{lemma}[\citet{fudenberg1991game}]\label{lemma:up}
The equilibrium set $\EQ(\theta)$ is upper-hemicontinuous in the belief $\theta$ if the utility function $u_i^s(\q)$ is continuous in $q$ for all $i \in I$ and all $s \in S$. 
\end{lemma}

\medskip 
\noindent\emph{Proof of Lemma \ref{lemma:constrained_set}.} 
\begin{itemize}
    \item[(i)] From Lemma \ref{lemma:up}, we know that such $\epsilon'$ must exist. 
    \item[(ii)] Since $\epsilonhat \leq \epsilon$, we know from Assumption \emph{(A3b)} that $F(\theta, \q) \subseteq N_{\delta}\(\EQ(\thetabar)\)$ for any $\theta \in N_{\epsilonhat}\(\thetabar\)$ and any $\q \in N_{\delta}\(\EQ(\thetabar)\)$. 
    \item[(iii)] Under Assumptions \ref{as:uh} -- \ref{as:convergence}, we know from Theorem \ref{theorem:convergence} that the sequence of the beliefs and strategies converges to a fixed point, denoted as $\(\theta^{\dagger}, \q^{\dagger}\)$ to be distinguished from the original fixed point $(\thetabar, \qbar)$.  If $\thetat \in N_{\epsilonhat}(\thetabar)$ for all $\t$, then $\lim_{\t \to \infty} \thetat = \theta^{\dagger} \in N_{\epsilonhat}(\thetabar) \subseteq N_{\thetaepfin}(\thetabar)$. Additionally, from \emph{(i)} and the fact that $\epsilonhat \leq \epsilon'$, we know that $\lim_{\t \to \infty} \qt = \q^{\dagger} \in \EQ(\theta^{\dagger})\subseteq  N_{\loadepfin}\(\EQ(\thetabar)\)$. Therefore, 
    \begin{align*}
        \lim_{\t \to \infty} \pro\(\thetat \in \neighinftheta(\thetabar) , ~ \qt \in  \neighinfload(\EQ(\thetabar))\)\geq &\pro\(\thetat \in  N_{\epsilonhat}(\thetabar), ~\forall \t\) \\
        \geq &\pro\(\thetat \in  N_{\epsilonhat}(\thetabar), \qt \in N_{\delta}(\EQ(\thetabar)), ~\forall \t\).
    \end{align*}
\end{itemize}
 \QEDA
 
 \vspace{0.2cm}
\noindent\emph{Proof of Proposition \ref{prop:global}.}
We first proof that \emph{(a)} is equivalent to \emph{(c)}. On one hand, if $\FP = \{\(\thetasran, \EQ(\thetasran)\)\}$, then for any initial state, the learning dynamics converges to a complete information fixed point with belief $\thetasran$ and strategy in $\EQ(\thetasran)$. That is, $\(\thetasran, \EQ(\thetasran)\)$ is globally stable. On the other hand, if there exists another fixed point $\(\theta^{\dagger}, \q^{\dagger}\) \in \FP \setminus \{\(\thetasran, \EQ(\thetasran)\)\}$, then learning that starts with the initial belief $\thetazero =\theta^{\dagger}$ (resp. $\thetazero =\thetasran$) and strategy $\q^1 = \q^{\dagger}$ (resp. $\q^1 = \qsran$) remains at $\(\theta^{\dagger}, \q^{\dagger}\)$ (resp. $\(\thetasran, \qsran\)$) for all stages w.p. 1. Thus, in this case, globally stable fixed points do not exist.

Next, we prove that \emph{(b)} is equivalent to \emph{(c)}. If $[\theta] \setminus \Sequiv(\q)$ is a non-empty set for any $\theta \in \Delta(\S)\setminus \{\thetasran\}$ and any $\q \in \EQ(\theta)$, then no belief with imperfect information $\theta \in \Delta(\S)\setminus \{\thetasran\}$ satisfies \eqref{eq:exclude_distinguished}. That is, only the complete information belief vector $\thetasran$ can be a fixed point belief. Therefore, all fixed point must be complete information fixed points. On the other hand, assume for the sake of contradiction that there exists a belief $\theta^{\dagger} \in \Delta(\S) \setminus \{\thetasran\}$ such that $[\theta^{\dagger}] \subseteq \Sequiv(\q^{\dagger})$ for an equilibrium strategy $\q^{\dagger} \in \EQ(\theta^{\dagger})$, then $\(\theta^{\dagger}, \q^{\dagger}\)$, which is not a complete information fixed point, is in the set $\FP$. Thus, we arrive at a contradiction. 

We can conclude that \emph{(a)} is equivalent to \emph{(b)} from the equivalence between \emph{(a)} and \emph{(c)}, and \emph{(b)} and \emph{(c)}. 
\QEDA

\medskip 
\noindent\emph{Proof of Proposition \ref{prop:complete_local}.}
First, since $\thetasran(\sran)=1$, $\thetasran$ satisfies the local consistency condition in \emph{(A3a)} for all $\q \in \Q$. Second, since $\Q$ is an invariant set for any strategy update and any $\theta \in \Delta(\S)$. We can choose an arbitrary $\epsilon$-neighborhood of $\thetasran$, and local invariance is satisfied. Therefore, neighborhood $(\Delta(\S), \Q)$ satisfies Assumption \ref{as:stability}. Thus, any complete information fixed point must be locally stable. \QEDA

\medskip 
\noindent\emph{Proof of Proposition \ref{prop:equivalent_complete}.}
From condition \emph{(i)} that $[\thetabar] \subseteq \Sequiv(\q)$ for any $\|\q- \qbar\| < \xi$, we have: 
\begin{align}\label{eq:expected}
    \mathbb{E}_{\thetabar}[u_i^{\s}(\q)] = u_i^{\sran}(\q), \quad \forall \i \in \I.
\end{align}
For any $\qbar \in \EQ(\thetabar)$, since $\qbar_i$ is a best response to $\qbar_{\mi}$ with respect to $\thetabar$, $\qbar_i$ must be a local maximizer of $\mathbb{E}_{\thetabar}[u_i^{\s}(\qi, \qbar_{\mi})]$. From \eqref{eq:expected}, $\qbar_i$ is a local maximizer of $u_i^{\sran}(\qi, \qbar_{\mi})$. From condition \emph{(ii)}, since the function $u_i^{\sran}(\qi, \qbar_{\mi})$ is concave in $\qi$, $\qbar_i$ is also a global maximizer of $u_i^{\sran}(\qi, \qbar_{\mi})$, and hence is a best response of $\qbar_{\mi}$ with complete information of $\sran$. Since this argument holds for all $\i \in \I$, $\qbar$ is a complete information equilibrium.\QEDA

 \medskip 
 \noindent\emph{Proof of Proposition \ref{prop:timescale}.} Since $\lim_{t \to \infty}\t_{t+1}-\kt = \infty$ with probability 1, as $t \rightarrow \infty$, the strategies between any two belief updates $\kt$ and $\t_{t+1}$ are updated with the static belief $\theta^{\kt}$. Under Assumption \ref{as:convergence}, we know that the sequence of strategies $\(q^j\)_{j=\kt}^{\t_{t+1}}$ converges to an equilibrium strategy in $\EQ(\theta^{\kt})$. From Lemma \ref{lemma:theta}, we know that $\lim_{t \to \infty} \theta^{\kt} = \thetabar$ with probability 1. Since the equilibrium set $\EQ(\theta)$ is upper hemicontinuous in $\theta$, the sequence of strategies must converge to an equilibrium $\qbar \in \EQ(\thetabar)$. We know from Lemma \ref{lemma:consistency} that the belief must form a consistent estimate of payoff distribution given $\qbar$. The proofs of local and global consistency are analogous to that in Theorem \ref{theorem:stability} and Proposition \ref{prop:global}, respectively, and thus are omitted. \QEDA




\end{document}